\documentclass[aps,prx,twocolumn,amsmath,amssymb,superscriptaddress,floatfix,reprint,longbibliography]{revtex4-2}
\usepackage{epsfig}
\usepackage{amsmath}
\usepackage{amssymb}
\usepackage{amsfonts}
\usepackage{mathptmx}
\usepackage{dcolumn}
\usepackage{eucal}
\usepackage{bm}
\usepackage{color}
\usepackage[colorlinks,linkcolor=blue,citecolor=blue]{hyperref}
\usepackage{epstopdf}
\usepackage{url}
\usepackage{svg}
\usepackage{braket}

\usepackage{array,makecell,tabularx,multirow}
\newcolumntype{C}[1]{>{\centering\let\newline\\\arraybackslash\hspace{0pt}}p{#1}}
\usepackage{hhline}
\usepackage{pifont}

\begin{document}
\title{Tunable room temperature nonlinear Hall effect \\ from the surfaces of elementary bismuth thin films}

\author{Pavlo~Makushko}
\affiliation{Helmholtz-Zentrum Dresden-Rossendorf e.V., Institute of Ion Beam Physics and Materials Research, 01328 Dresden, Germany}

\author{Sergey~Kovalev}
\affiliation{Helmholtz-Zentrum Dresden-Rossendorf e.V., Institute of Radiation Physics, 01328 Dresden, Germany}

\author{Yevhen~Zabila}
\affiliation{Helmholtz-Zentrum Dresden-Rossendorf e.V., Institute of Ion Beam Physics and Materials Research, 01328 Dresden, Germany}
\affiliation{The H. Niewodniczanski Institute of Nuclear Physics,
Polish Academy of Sciences, 31–342 Krakow, Poland}

\author{Igor~Ilyakov}
\affiliation{Helmholtz-Zentrum Dresden-Rossendorf e.V., Institute of Radiation Physics, 01328 Dresden, Germany}

\author{Alexey~Ponomaryov}
\affiliation{Helmholtz-Zentrum Dresden-Rossendorf e.V., Institute of Radiation Physics, 01328 Dresden, Germany}

\author{Atiqa~Arshad}
\affiliation{Helmholtz-Zentrum Dresden-Rossendorf e.V., Institute of Radiation Physics, 01328 Dresden, Germany} 

\author{Gulloo~Lal~Prajapati}
\affiliation{Helmholtz-Zentrum Dresden-Rossendorf e.V., Institute of Radiation Physics, 01328 Dresden, Germany} 

\author{Thales~V.~A.~G.~de~Oliveira}
\affiliation{Helmholtz-Zentrum Dresden-Rossendorf e.V., Institute of Radiation Physics, 01328 Dresden, Germany} 

\author{Jan-Christoph~Deinert}
\affiliation{Helmholtz-Zentrum Dresden-Rossendorf e.V., Institute of Radiation Physics, 01328 Dresden, Germany} 

\author{Paul~Chekhonin}
\affiliation{Helmholtz-Zentrum Dresden-Rossendorf e.V., Institute of Ion Beam Physics and Materials Research, 01328 Dresden, Germany}
\affiliation{Helmholtz-Zentrum Dresden-Rossendorf e.V., Institute of Resource Ecology, 01328 Dresden, Germany}

\author{Igor~Veremchuk}
\affiliation{Helmholtz-Zentrum Dresden-Rossendorf e.V., Institute of Ion Beam Physics and Materials Research, 01328 Dresden, Germany}

\author{Tobias~Kosub}
\affiliation{Helmholtz-Zentrum Dresden-Rossendorf e.V., Institute of Ion Beam Physics and Materials Research, 01328 Dresden, Germany}

\author{Yurii~Skourski}
\affiliation{Helmholtz-Zentrum Dresden-Rossendorf e.V., Dresden High Magnetic Field Laboratory (HLD-EMFL), 01328 Dresden, Germany}

\author{Fabian~Ganss}
\affiliation{Helmholtz-Zentrum Dresden-Rossendorf e.V., Institute of Ion Beam Physics and Materials Research, 01328 Dresden, Germany}

\author{Denys~Makarov}
\email{d.makarov@hzdr.de}
\affiliation{Helmholtz-Zentrum Dresden-Rossendorf e.V., Institute of Ion Beam Physics and Materials Research, 01328 Dresden, Germany}

\author{Carmine~Ortix}
\email{cortix@unisa.it}
\affiliation{Dipartimento di Fisica ``E. R. Caianiello", Universit\`a di Salerno, IT-84084 Fisciano (SA), Italy}

\maketitle

\textbf{The nonlinear Hall effect (NLHE) with time-reversal symmetry constitutes the appearance of a transverse voltage quadratic in the applied electric field. It is a second-order electronic transport phenomenon that induces frequency doubling and occurs in non-centrosymmetric crystals
with large Berry curvature -- an emergent magnetic field encoding the geometric properties of electronic wavefunctions. 
The design of (opto)electronic devices
based on the NLHE is however hindered by the fact that this nonlinear effect 
typically appears at low temperatures and in complex compounds characterized by Dirac
or Weyl electrons
Here, we show a strong room temperature NLHE in the centrosymmetric elemental material bismuth synthesized in the form of technologically relevant polycrystalline thin films.
The ($1\,1\,1$) surface electrons of this material are equipped with a Berry curvature triple
that activates side jumps and skew scatterings generating nonlinear transverse currents.
We also report a boost of the zero field nonlinear transverse voltage in arc-shaped bismuth stripes due to a
extrinsic geometric classical counterpart of the NLHE
This electrical frequency doubling in curved geometries is then extended to optical second harmonic generation in the terahertz (THz) spectral range.
The strong nonlinear electrodynamical responses of the surface states 
are further demonstrated
by 
a concomitant
highly efficient 
THz
third harmonic generation which we achieve in a broad range of frequencies in Bi and Bi-based heterostructures. 
Combined with the possibility of growth on CMOS-compatible and mechanically flexible substrates, these 
results highlight the potential of Bi thin films for THz (opto)electronic applications. }

Second-order electrical responses are fundamentally prohibited in material structures with inversion symmetry. Essential nonlinear electrodynamical operations such as frequency mixing and current rectification are thus sought in non-centrosymmetric materials. This excludes the largest class of materials known to mankind: non-magnetic ones with bulk inversion symmetry. However, this paradigm is strictly valid only in bulk three-dimensional materials. At surfaces and interfaces, the centrosymmetry of the structure is naturally broken, and certain second harmonic electrodynamical processes are symmetry-allowed. For instance, the nonlinear Hall effect (NLHE)~\cite{Sodemann2015,Ortix2021,du21} due to the Berry curvature dipole can be non-vanishing at surfaces with unusually low crystalline symmetries~\cite{waw22}.  
Surface-induced nonlinear electrical processes have been also observed~\cite{he21} at the (111) surface of Bi$_2$Se$_3$ -- a strong three-dimensional topological insulator with trigonal symmetry~\cite{Zhang2009}. 
The threefold rotational symmetry of its crystal forces 
the Berry curvature dipole associated with the topological surface Dirac cone to vanish. However, a higher order moment, dubbed Berry curvature triple (BCT), activates 
side jumps \cite{du19,xia19}
and skew scatterings
capable of inducing 
large nonlinear 
responses. 
This effect, dubbed as quantum frequency doubling in Ref.\cite{he21} and disorder-induced NLHE in Ref.~\cite{du21}, is
different with respect to 
the nonlinear Hall effect related to
the Berry curvature dipole only in that 
the transversal nonlinear currents are related to
dissipative longitudinal responses [see Ref.~\cite{Ortix2021} and Supplementary Information].
In the remainder, we will adopt the nomenclature of Ref.~\cite{du21}, and refer to this effect as NLHE capitalizing on the fact that what is observed is an electrical second-harmonic generation in the transverse channel related to the geometric properties of the wavefunctions.

Second harmonic generation (SHG) and current rectification due to geometric properties of Bloch electrons~\cite{ore21} have been predicted to be highly efficient at 
high
frequencies and thus possibly relevant for optoelectronic applications~\cite{zha21,iso20} in the terahertz spectral domain.
To bring this concept to reality, a number of criteria should be fulfilled [\textit{c.f.} Table~\ref{table1}]. 
First, the nonlinear transverse responses should be sizable at room temperature and above. Second, the NLHE voltage should be controllable by a non-thermal parameter and thus be tunable. Third, NLHE-based devices should rely on technologically relevant material fabrication, 
for instance, with the use of CMOS-compatible ({\it e.g.} silicon) and mechanically flexible substrates ({\it e.g.} polymers).
This can be achieved by first searching for materials with a simplified chemical composition, ideally single-element systems, and then for low-cost microstructures: polycrystalline thin films rather than single crystals. 
Finally, toxic heavy rare-earth materials should be eliminated all together, thus paving a way towards green 
(opto)electronics. 
Room-temperature nonlinear Hall effect has been observed in the ternary type-II Weyl semimetal~\cite{kum21} TaIrTe$_4$
and in the Dirac material 
BaMnSb$_2$
~\cite{min23}. Tunability via gating has been instead theoretically predicted and experimentally verified at low temperatures in low-dimensional materials including transition metal dichalcogenides~\cite{xu18,you18,zha18,du18,son19,ma19,kan19,sin20,ma22}, bilayer graphene~\cite{bat19,ho21}, and oxide interfaces~\cite{les22}. 
Here, we report the discovery of the first material structure fulfilling all the technologically relevant criteria at the same time: (111) polycrystalline thin films of elemental bismuth.
 The strong spin-orbit coupled surface states of this element~\cite{kor08,du16} are equipped with a finite Berry curvature that averages to zero but possesses a finite BCT. 
Importantly, the room-temperature NLHE of this green~\cite{moh10} semimetal can be efficiently tuned using a curvature-induced classical shape effect~\cite{gen22}, namely the geometric NLHE whose signatures have been reported so far only in graphene~\cite{sch19}. 
We show that this effect is present in full force 
in polycrystalline thin films and cooperates with the BCT to create second harmonic transverse electrical responses. 
Moreover, we report a complementary optical SHG at THz frequencies in arrays of arc-shaped bismuth stripes, thereby providing for the first time signatures of the connection between nonlinear transport at low frequencies and optical nonlinearities at higher frequencies.
 
\begin{figure*}
\centering
\includegraphics[width=0.95\textwidth]{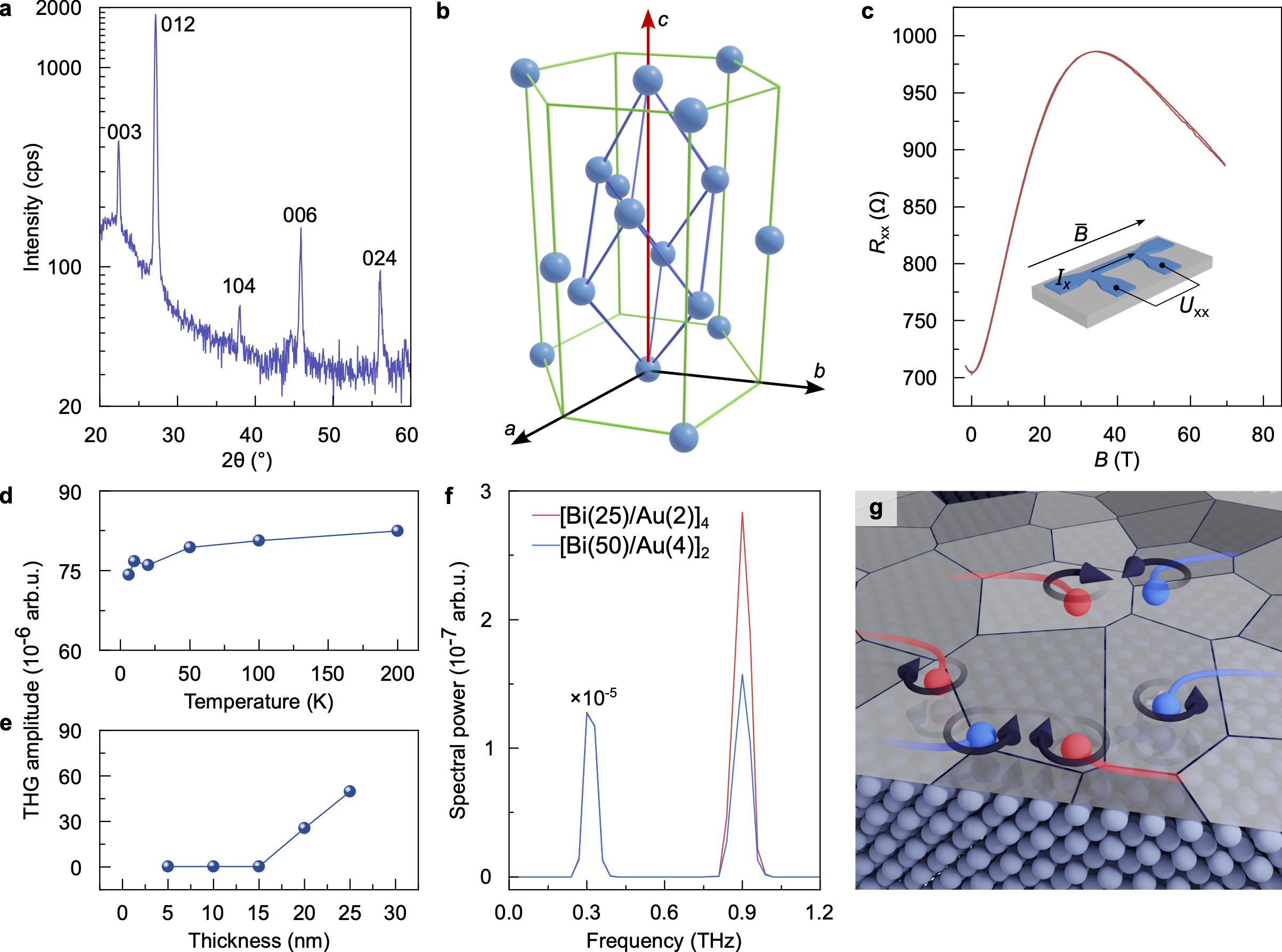}
    \caption{\textbf{Quantum properties and surface states of polycrystalline Bi thin films.} 
    (a) XRD data confirming the ($0\,0\,1$) [hexagonal] // ($1\,1\,1$) [rhombohedral] texture of the investigated bismuth thin film (see also Supplementary Information). 
    (b) Unit cell of elemental bismuth in hexagonal (green lines) and rhombohedral (blue lines) representation (not all atoms are shown). Red line represents the three-fold rotation symmetry axis.
    (c) High-field magnetoresistance data. An inset schematic shows the device under test and the direction of the applied magnetic field. 
    (d) Temperature dependence of the third harmonic generated (THG) signal amplitude. Varying temperature from $4$\,K up to room temperature there is a less than $10$\,\% variation that can be ascribed to temperature drifts of the experimental setup. 
    (e) bismuth film thickness dependence of the third harmonic generated (THG) signal amplitude, indicating that THG in Bi thin films originates from surface states. 
    (f) Power spectrum of the THz THG measured of Bi/Au heterostructures. 
    (g) Schematic of side jumps and skew scattering. Self-rotating surface Bloch electrons are deflected like the Magnus effect. }
\label{fig:fig1}
\end{figure*}

We prepared Bi thin films with a nominal thickness of $100$~nm by magnetron sputtering at room temperature. The thin films are characterized using X-ray diffraction (XRD) [\textit{c.f.} Fig.~\ref{fig:fig1}(a)], which reveals a polycrystalline microstructure with grains of typical size of $\simeq 33$~nm. 
Furthermore, XRD and electron backscatter diffraction (EBSD) data [Supplementary Information] indicate a preferred rhombohedral (111) crystallographic texture [\textit{c.f.} Fig.~\ref{fig:fig1}(b)], and hence the possibility of non-trivial geometric properties of the surface electronic waves in the material~\cite{bat21}. The EBSD pole figures indicate the thin films to be structurally isotropic within the film plane. 

The polycrystalline nature of the Bi thin films does not hinder the appearance of quantum phenomena conventionally observed in single crystals. This is verified in Fig.~\ref{fig:fig1}(c), where we show that the high-field magnetoresistance is characterized by a drop when the externally applied in-plane magnetic field $B>35$~T. This is consistent with a previous study~\cite{zhu17} that related such high-field magnetoresistance drop with the complete drying of a Fermi sea of electrons confined to their lowest Landau levels, and thus well inside the so-called quantum limit~\cite{abr98} attained in Bi crystals at magnetic fields $B>15$~T.

The high-field magnetotransport measurement of Fig.~\ref{fig:fig1}(c) probes a quantum response 
associated with the bulk hole and electron states. However, it makes no assertion on the presence and physical properties of surface states. 
To overcome this hurdle, we employ a terahertz third harmonic generation (THG) approach, which was previously used to study the Dirac states at the surface of topological insulators~\cite{kov21}. Specifically, a THz excitation at the fundamental frequency of 300\,GHz is applied with normal incident pulse. 
The generated third harmonic at 900\,GHz is characterized in time-domain using electro-optical sampling. 
A systematic analysis of the THz THG in our Bi thin films proves the surface origin of the signal, analogously to the surface-mediated THz nonlinearities in topological insulators~\cite{gio16}. 
The THz THG amplitude  
reveals only very weak temperature dependence, 
as shown in Fig.~\ref{fig:fig1}(d). 
We also observe a quenching of the THz THG signal for Bi films thinner than 15\,nm [\textit{c.f.} Fig.~\ref{fig:fig1}(e) and Supplementary Information]. This 
indicates a surface band gap opening due to the hybridization of surface states belonging to opposite surfaces.  
Further evidence that high harmonic generation in Bi is predominantly due to its surface states comes from the study of Bi/Au multilayers. As shown in Fig.~\ref{fig:fig1}(f), a stack of four Bi(25\,nm)/Au(2\,nm) heterolayers has a more efficient THz THG as compared to two stacked Bi(50\,nm)/Au(4\,nm) bilayers, in agreement with the fact that in the former multilayer the number of surfaces is doubled. The linear THz emission is instead the same for the two multilayers since the total thickness of the Bi and Au layers are equal. 

The observed highly nonlinear kinetics at THz frequencies of the Bi surface states can be ascribed to a low surface carrier density -- this implies a reduced heat capacity and thus enhanced thermodynamic nonlinearity~\cite{kov21SA} -- and to surface band dispersions that contain terms growing linearly with the crystalline momentum~\cite{kov20}. 
Bismuth has been recently classified as an higher-order topological insulator~\cite{schind18}, with a trivial value of the ${\mathbb Z}_2$ topological invariant~\cite{fu07} for systems in the AII class of the Altland-Zirnbauer classification~\cite{alt97}. This immediately implies that the surface states cannot realize Dirac cones [Supplementary Information]. Nevertheless, the strong spin-orbit coupling of Bi combined with the surface inversion symmetry breaking yields a sizable, linear in $k$, Rashba spin-orbit coupling. 
Additionally, the (111) texture of our polycrystalline samples leads to a hexagonal warping that has a twofold effect. 
First, the surface bands have Fermi lines with a hexagonal ``snowflake" shape~\cite{fu09}. Second, electrons are equipped with a non-vanishing local Berry curvature driven entirely by crystalline anisotropy effects. 
These two properties enable 
both nonlinear side-jumps~\cite{du19,Ortix2021,xia19}
and nonlinear skew scattering 
due to the fact that surface Bloch electrons with opposite velocities self-rotate in opposite directions and give rise to a net Magnus effect [\textit{c.f.} Fig.~\ref{fig:fig1}(g)] quantified by the BCT [Supplementary Information]. Importantly, 
the third-order skew scattering deflections 
of surface electrons belonging to different Rashba branches do not cancel each other and lead to a net 
contribution to the
NLHE [Supplementary Information]. 

\begin{figure*}
\centering
 \includegraphics[width=0.95\textwidth]{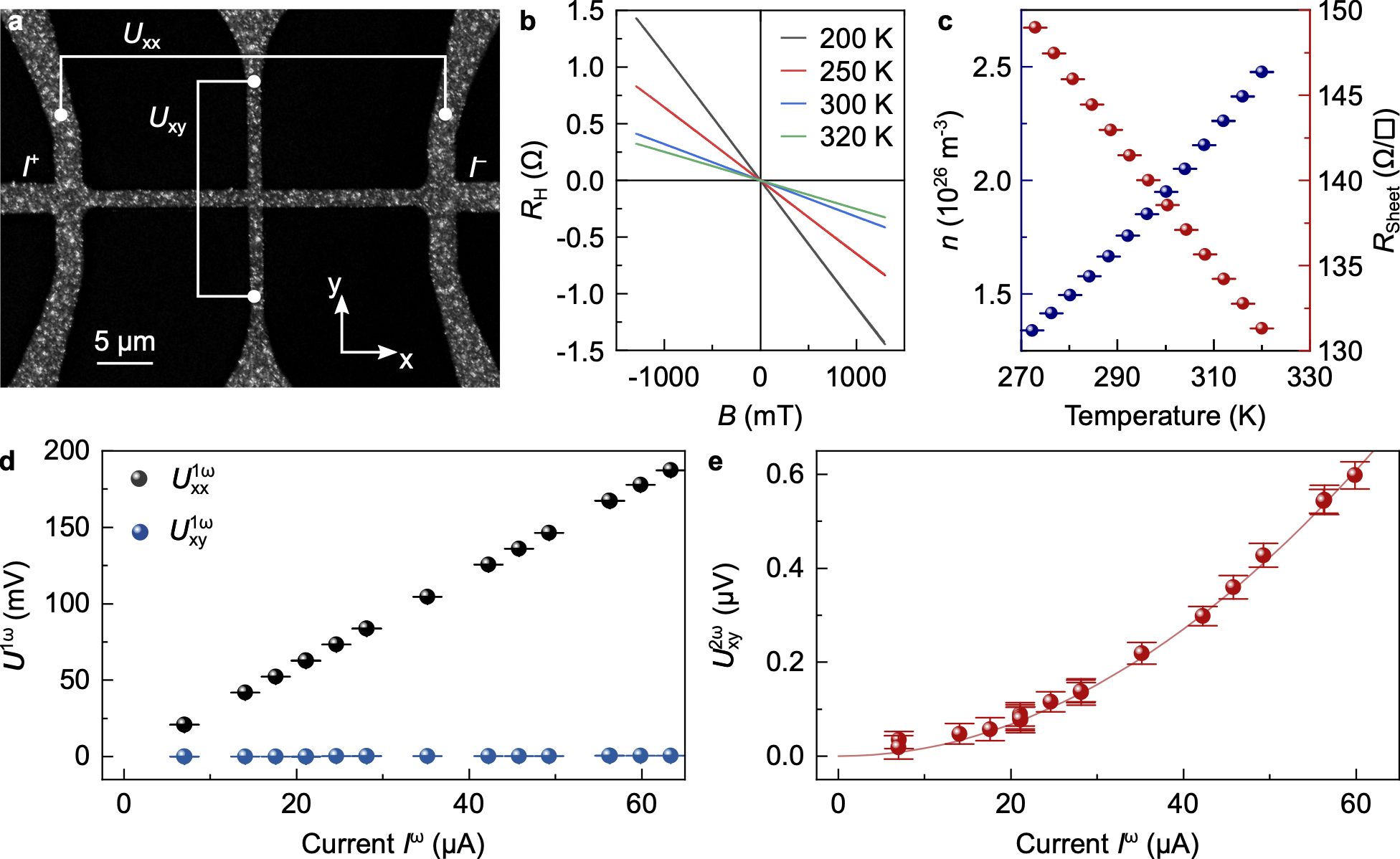}
	\caption{\textbf{Magnetotransport and nonlinear Hall effect in Bi thin films.} 
 (a) Electron microscopy image of a Hall cross device and schematic configuration of the transverse harmonic measurement. 
 (b) Hall resistance $R_\text{H}$ measured of an extended 100-nm-thick Bi film at selected temperatures. 
 (c) Temperature evolution of the charge carriers density $n$ and sheet resistance $R_\text{Sheet}$ of the extended bismuth thin film. 
 (d) First harmonic longitudinal $U_\text{xx}^{1\omega}$ and transverse $U_\text{xy}^{1\omega}$ voltages vs current amplitude measured of the Hall cross structures shown in panel (a). 
 (e) Second harmonic transverse voltage $U_\text{xy}^{2\omega}$ vs current amplitude. Symbols correspond to the measured data and solid line is a quadratic fit $R_\text{yxx}^{2\omega} \cdot I^2$, where $R_\text{yxx}^{2\omega}$ is the nonlinear transverse resistance. Harmonic transport measurements are carried out at the fundamental frequency of 787\,Hz. }
	\label{fig:fig2}
\end{figure*}

To verify the occurrence of this quantum transport effect, we pattern our polycrystalline samples in a Hall cross geometry [\textit{c.f.} Fig.~\ref{fig:fig2}(a)]. We first perform temperature dependent Hall resistance measurements by applying an external out-of-plane magnetic field. Consistently with the high-field magnetotransport characterization, we find Hall transport to be dominated by bulk carriers: the Hall coefficient displays a strong temperature dependence [\textit{c.f.} Fig.~\ref{fig:fig2}(b)]. The associated bulk carrier density of $n = 2\cdot10^{26}$\,m$^{-3}$ at room temperature [\textit{c.f.} Fig.~\ref{fig:fig2}(c)] is consistent with previous reports on Bi thin films 
and single crystals. 
This holds true also when considering the temperature decrease of the sheet resistance [\textit{c.f.} Fig.~\ref{fig:fig2}(c)]. 

Zero-field transport properties are obtained by sourcing an oscillating current with frequency $\omega/(2 \pi)$ along the $\hat{x}$ direction, and concomitantly measuring the longitudinal $U_{xx}^{\omega}$ as well as the first ($U_{xy}^{\omega}$) and second ($U_{xy}^{2\omega}$) harmonic transverse voltages in a conventional lock-in detection scheme. 
Fig.~\ref{fig:fig2}(d) [see also Supplementary Information] shows the ensuing current-voltage characteristic of the first harmonic longitudinal response, which 
remains linear
over the full range of sourced currents [Supplementary Information]. This Ohmic behavior allows us to exclude a leading role played by Schottky barriers. 
Furthermore, using a bimaterial model, we estimate 
thermal effects due to Joule heating 
to be negligible [Supplementary Information].
We also find the first harmonic transverse voltage $U_{xy}^{\omega}$ [\textit{c.f.} Fig.~\ref{fig:fig2}(d)] to be three order of magnitudes smaller than $U_{xx}^{\omega}$. This is consistent with the fact that in time-reversal symmetric conditions linear transverse responses are symmetry-allowed only in crystals of sufficiently low symmetry. 
The threefold rotation symmetry of Bi implies that the in-plane resistivity can be described by a single scalar. Therefore, the transverse voltage $U_{xy}^{\omega}$ is forced to vanish both in single crystals and in polycrystalline (111) Bi thin films. 

Second-order transverse responses are instead completely compatible with a threefold rotational symmetry [Supplementary Information]. The bulk centrosymmetry of Bi however implies that any second-order electrical response, if finite, must originate entirely from its surface states. 
Fig.~\ref{fig:fig2}(e) demonstrates that our polycrystalline Bi thin films possess a surface-induced second-order transverse response. 
We note that the transverse voltage $U_{xy}^{2\omega}$ has a characteristic quadratic current-voltage behaviour: 
the measured signal can be reliably fitted by the relation $U_{xy}^{2\omega}(I_x) = R_\text{yxx}^{2\omega} \cdot I_x^2$ with $R_\text{yxx}^{2\omega} = 166$\,$\Omega$/A.
Combined with the fact that the response occurs at double the driving frequency, this establishes the electronic second-order nature of the effect. 
We show this nonlinear transverse response to be present also in a Bi device grown on mechanically flexible polymeric substrate [Supplementary Information].
Furthermore, the surface-mediated NLHE of Bi establishes a general approach to generate this effect in other elemental materials. Various noble metals, such as Pt, Rh or Ir, do feature surface states \cite{waw22}. As proof of concept, we verify the presence of a nonlinear Hall voltage in 5-nm-thick Pt thin film [Supplementary Information].

Being related to an intrinsic geometric property of surfaces electronic wavefunctions in a bulk material, the electronic second harmonic generation due to the BCT cannot be tuned using a non-thermal control parameter. This is different from the gate dependence widely reported at low temperatures in low-dimensional materials with finite Berry curvature dipoles~\cite{xu18,ma19,ho21,les22}. 
Nevertheless, we  are able to reach control of the magnitude of the nonlinear transverse voltage 
using a curvature-induced classical shape effect~\cite{gen22}: the so-called geometric nonlinear Hall effect~\cite{sch19}. 
This phenomenon is due to the spontaneous appearance in bent conducting channels of a transverse electric potential, which is necessary to accelerate radially the charge carriers and let them follow the curved path of the material microstructure. 
As a result, surface charges similar in nature to those of the classical Hall effect arise, even in the complete absence of external perpendicular magnetic fields. The presence of time-reversal symmetry implies that the electrical transverse potential needs to be quadratic in the injected current density, qualifying this curvature-induced effect as nonlinear Hall. Importantly, the loss of centrosymmetry necessary for the occurrence of the nonlinear electrical response is obtained at the level of the mesoscale geometry of the material structure, in strict analogy with the flexoelectric effect~\cite{zub13}. Conceptually speaking, a geometric nonlinear Hall effect is therefore allowed even in centrosymmetric crystals. In the specific case of bismuth thin films, this implies that both surface charge carriers and bulk charge carriers can in principle give rise to a geometric nonlinear Hall response. 

\begin{figure*}
\centering
 \includegraphics[width=0.95\textwidth]{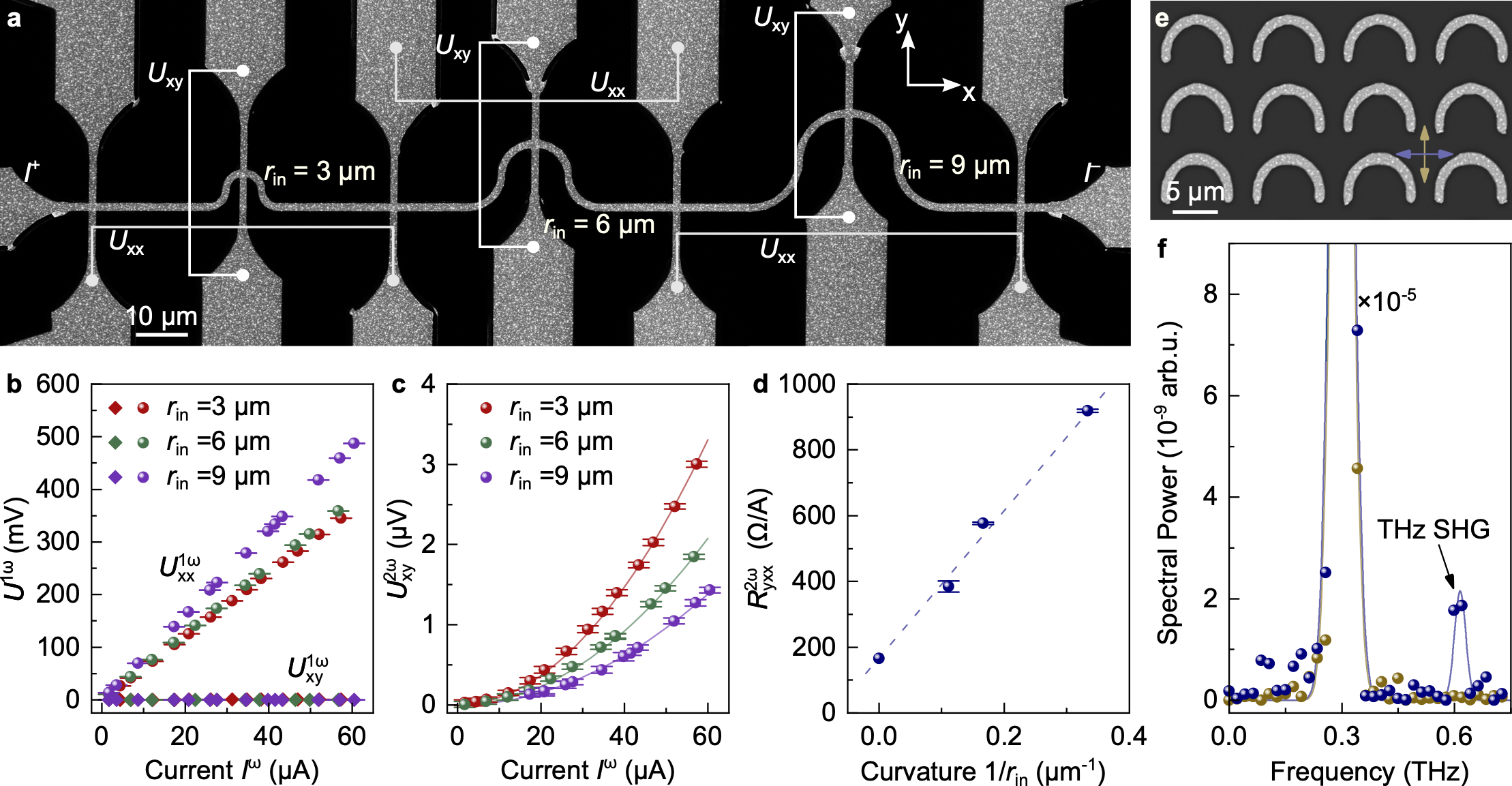}
	\caption{\textbf{Geometric nonlinear Hall effect in arc-shaped bismuth Hall bars.} 
 (a) Electron microscopy image of the Hall bar device with curved longitudinal channels of different inner radii ($R_\text{in} = 3\,\mu$m, $6\,\mu$m, $9\,\mu$m) and width $1.4\,\mu$m. The schematics show the configuration of transverse harmonic measurement at each arc-shaped Hall bar. 
 (b) First harmonic longitudinal $U_\text{xx}^{1\omega}$ and transverse $U_\text{xy}^{1\omega}$ voltages vs current amplitude measured at each of arc-shaped Halls bars. 
 (c) Second harmonic transverse $U_\text{xy}^{2\omega}$ voltages vs current amplitude. Symbols correspond to the measured data and solid lines are quadratic fit $R_\text{yxx}^{2\omega} \cdot I^2$.
 Harmonic transport measurements are carried out at the fundamental frequency of $787$\,Hz.
 (d) Scaling of the nonlinear transverse resistance $R_\text{yxx}^{2\omega}$ with the inverse inner radius $R_\text{in}$ of the arc structures. Zero curvature corresponds to the data of the bismuth cross shown in Fig.~\ref{fig:fig2}(e). Dashed line is guide to the eye.
 (e) Electron microscopy image of an array of $100$-nm-thick bismuth arcs (inner radius: $3\,\mu$m; width: $1.2\,\mu$m) used for THz second harmonic generation (SHG) measurement. Arrows indicates relative polarization of the incident THz radiation.
 (f) Transmitted spectral power of the THz signal measured of the sample shown in panel (e). A peak at $0.6\,$THz indicates generation of the second harmonic signal (fundamental THz frequency is $0.3\,$THz). The color of the lines in panel (f) corresponds to the color of the arrows in panel (e), indicating the polarization of the incident THz pulse. Intensity of the fundamental beam is suppressed by bandpass filters located between the sample and the detector. The background for each curve is also suppressed by
subtracting measurements done in antiparallel polarization of THz pulse relative to the symmetry axis of the arcs, \textit{i.e.} $0-180^\circ$ and $90-270^\circ$.} 
	\label{fig:fig3}
\end{figure*}

We verify the occurrence of this geometric effect by patterning arc-shaped Hall bars [\textit{c.f.} Fig.~\ref{fig:fig3}(a)] with the longitudinal arc-shaped channels that have inner curvature 
radii,
$r_{in}$, of $\simeq 3,\,6$ and $9\,\mu$m and a width in the radial direction of $w \simeq 1.2\,\mu$m: this quasi-one-dimensional geometry of the longitudinal channel guarantees substantial geometry-induced modifications of charge motion. We first verify [\textit{c.f.} Fig.~\ref{fig:fig3}(b)] that electrical transport in the linear response regime is Ohmic with the current-voltage characteristic that is 
linear
in all the range of sourced currents.  
The corresponding longitudinal resistivities are  independent of the specific curvature radius and similar
to the conventional Hall cross devices discussed in Fig.~\ref{fig:fig2}(c). 
Fig.~\ref{fig:fig3}(b) also shows that the transverse voltage is suppressed indicating the absence of geometric effects in the linear response [see Supplementary Information]. 

The nonlinear response has instead a completely different behavior. As shown in Fig.~\ref{fig:fig3}(c), the second harmonic transverse voltage of the arc with inner radius $r_{in}=3\,\mu$m inner radius has about $500\,\%$ increase with respect to the Hall cross. 
The geometric NLHE of bismuth 
can be larger than 
the contribution from the BCT of its surface states. 
Our data show that control of 
the mesoscale curvature of the conducting channel can be used to 
tune
the nonlinear transverse voltage $U_{xy}^{2\omega}$. 
This is in agreement with the fact that the
geometric contribution  
is expected to scale
inversely with the inner arc radius $r_{in}$ as~\cite{sch19}
$$U_{xy,\, geo}^{2 \omega} \simeq \dfrac{1}{r_{in} w t^2} \, \dfrac{m}{n^2 q^3} \, I_x^2,$$
where 
$t,w$ indicate the thickness and width of the conducting channel, respectively, while
$n,m,q$ indicate the density, mass and charge of the carriers, respectively. 
Figure~\ref{fig:fig3}(d) shows that the Bi arcs follow the curvature radius scaling of the nonlinear Hall voltage predicted in the equation above.
We can also attribute the sizable geometric NLHE of the arc-shaped Hall crosses predominantly to the Bi surface states. Considering the charge carrier density obtained from magnetotransport measurement [Fig.~\ref{fig:fig2}(c)], the bulk contribution to $U_{xy, \, geo}^{2 \omega}$ lies in the hundreds of pV for a 100~nm-thick film.
When considering the surface states contribution, the film thickness has to be replaced with their nanometer sized penetration depth. This can lead to a strong boost of the transverse voltage up to the scale of $\mu$V in agreement with our experimental observations.  

The strong enhancement of the NLHE due to curvature effects at low frequencies is complemented by the occurrence of  optical second harmonic generation (SHG) in the terahertz spectral domain. Extended arrays of $100$-nm-thick Bi arcs with inner radius of $3\,\mu$m [see Fig.~\ref{fig:fig3}(e)] display a SHG signal at $600$\,GHz in response to an $300$\,GHz excitation. We also observe a strong polarization dependence in the THz SHG signal of Bi arcs. As shown in Fig.~\ref{fig:fig3}(f) the second harmonic peak is enhanced when the incident electric field is polarized along the arc, \textit{i.e.} perpendicular to the symmetry axis of an individual arc structure. This is consistent with the main feature of the geometric NLHE.
Furthermore, the universality of the geometric nonlinear transport effect is paralleled by the ubiquitous occurrence of THz SHG in devices with curved geometries. As shown in  Supplementary Information, arrays of 20-nm-thick Au arcs display an even stronger SHG signal, even though THz THG is completely absent in Au thin films.
In Ref.~\cite{wen23}, SHG was observed in devices with gold split-ring resonators arrays and attributed to the nonlinear Thomson scattering. We note that in our devices this mechanism is unlikely to be responsible of the SHG at the fundamental frequency of $0.3$\,THz  due to the small dimensions of Au and Bi arcs. We can therefore attribute the THz SHG of our study to geometric effects related to Bi and Au surface states. Intrinsic second-order nonlinear effects related to well-known surface states of Au~\cite{yan15} have been anticipated in Ref.~\cite{wen23}.

\begin{figure*}
\centering
 \includegraphics[width=0.95\textwidth]{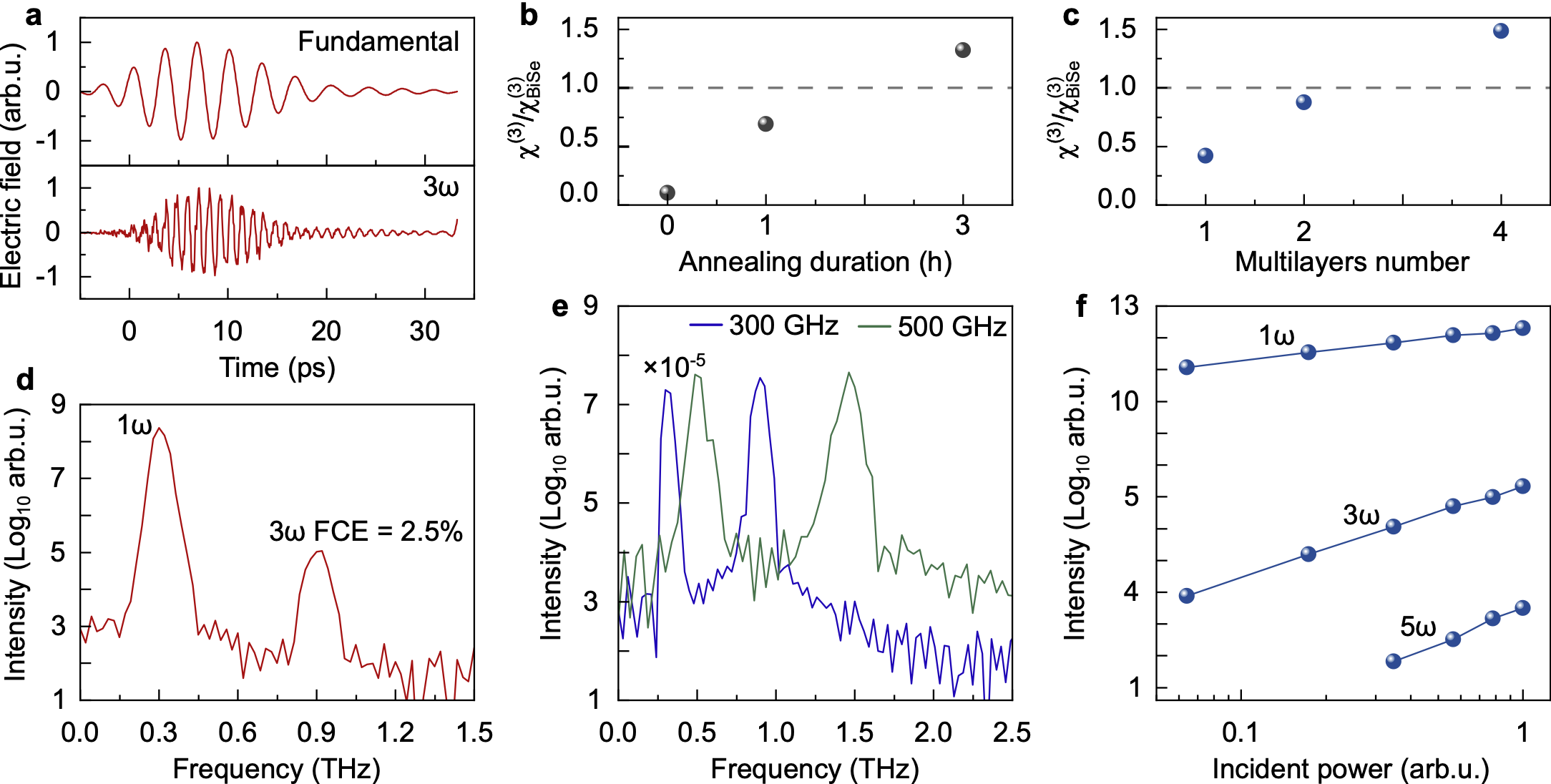}
	\caption{\textbf{Highly-efficient and tunable THz high harmonic generation in bismuth-based systems.} 
 (a) Time-domain dynamics of the THz light at fundamental frequency $\omega = 300\,$GHz (top panel) and generated $3\omega$ signal (bottom panel) measured through electro-optic sampling taken after Bi(100\,nm)/Au(2\,nm) thin film was annealed at 200$^{\circ}$C for 3\,h.
 (b)  Tailoring the efficiency of the THz THG in Bi(100\,nm)/Au(2\,nm) thin films by thermal annealing in vacuum at 200$^{\circ}$C. 
  (c) Tailoring the efficiency of the THz THG in Bi/Au heterostructures: \#1 = Bi(100\,nm)/Au(2\,nm), \#2 = [Bi(50\,nm)/Au(4\,nm)]$_2$, \#4 = [Bi(25\,nm)/Au(2\,nm)]$_4$.
 (d) Transmitted spectral power of the THz signal taken from the Bi(100\,nm)/Au(2\,nm) thin film annealed at 200$^{\circ}$C for 3\,h. The corresponding time-domain data is shown in panel (a). Here, no band-pass filter was applied during measurement. The data reveal high field conversion efficiency (FCE) of the THz THG at $2.5\,\%$.
 (e)  Intensity of the transmitted signal taken of the Bi(100\,nm)/Au(2\,nm) thin film exposed to the THz light at fundamental frequency of 300\,GHz and 500\,GHz.
  (f) Spectral power of the 1$^\text{st}$, 3$^\text{rd}$ and 5$^\text{th}$ harmonics measured in [Bi(25\,nm)/Au(2\,nm)]$_4$ heterostructure. The fundamental frequency of the incident light is 300\,GHz.
In panels (b) and (c), the plots show the third-order susceptibility $\chi^{(3)}$ measured in Bi samples normalized to the efficiency of the THz THG in epitaxially grown 100-nm-thick Bi$_2$Se$_3$ films~\cite{tie22}.
 }
	\label{fig:fig4}
\end{figure*}

The room temperature nonlinear 
electrodynamic
responses featured above indicate that elemental bismuth 
generally
possess 
large THz nonlinear susceptibilities and high harmonic conversion efficiencies. 
We observe [\textit{c.f.} Fig.~\ref{fig:fig4}(a)] highly-efficient room temperature THz THG with the efficiency that can be tuned changing the thin film microstructure either via thermal processing [\textit{c.f.} Fig.~\ref{fig:fig4}(b) and Supplementary Information] or by growing Bi/Au heterolayers [\textit{c.f.} Fig.~\ref{fig:fig1}(f), Fig.~\ref{fig:fig4}(c) and Supplementary Information]. 
We find a field conversion efficiency (FCE) of the third harmonic that can reach values $\simeq 2.5\%$ [\textit{c.f.} Fig.~\ref{fig:fig4}(d)] thus surpassing the performance of Bi$_2$Se$_3$: the benchmark material providing by now the highest known efficiency of THz THG~\cite{tie22}.
Note that for the nonlinear transport in the DC limit  Bi$_2$Se$_3$ has been reported~\cite{he21} to display nonlinear Hall voltages that reach a few $\mu$V sourcing currents of 1\,mA. In  Bi polycrystalline thin films, similar nonlinear Hall voltages are reached [\textit{c.f.} Fig.~\ref{fig:fig3}(c)] sourcing currents of 40\,$\mu$A. This tenfold decrease in the required current and the fact that the DC NLHE in Bi$_2$Se$_3$ occurs only at temperatures $<200$~K additionally demonstrate the advantages of Bi thin films for nonlinear electrical and electrodynamic responses. 

We also demonstrate efficient THz THG in a broad frequency range [\textit{c.f.} Fig.~\ref{fig:fig4}(e)]. Upon increase of the fundamental frequency from 300\,GHz to 500\,GHz, high efficiency of the generated third harmonic retains even at 1.5\,THz. Furthermore, the fluence dependence of THz THG and fifth harmonic does not show any saturation [\textit{c.f.} Fig.~\ref{fig:fig4}(f)], see also [Supplementary Information]. 
This indicates the absence of detrimental heat accumulations effects, and thus of ultrafast electronic heat dissipation due to Coulomb scattering between surface-state charges and bulk charges~\cite{pri22}. 
The ensuing possibility of using pulses of longer duration enabling parametric processes together with the observed broad-range high efficiency paves the way to green optoelectronic devices for prospective communication technologies, \textit{e.g.} sixth generation (6G) networks. In this regard, elementary bismuth offers key advantages due to established commodity scale production, commercial availability and possibility of straightforward processing as technologically relevant thin films. Furthermore, due to its notorious magnetoresistance and Hall effects~\cite{yan99}, bismuth is also relevant for sensor applications~\cite{mel15,mat22}. Finally, the combination of nonlinear electrodynamical capabilities and higher-order topological properties leading to spin-momentum locked hinge modes can be used to integrate optoelectronic and spintronic functionalities in a single elemental material system. 
Topological properties have been recently shown to appear even in amorphous matter~\cite{cor23}. 
From a fundamental point of view, our results represent the first step forward in realizing Berry curvature-mediated effects in polycrystalline structures, 
and thus open a vast space of exploration of quantum effects related to the electronic wavefunction geometry using modern thin film technology.

\begin{table*}
    \caption{Materials for the nonlinear Hall effect {\it vs} relevant criteria for their technological exploitation.}
    \label{table1}
    \begin{tabular}{C{50mm}||c|c|c|c}
         \hhline{=:t:====}
         Materials for nonlinear Hall & Chemical composition & Operating temperature & Tunability & Green \\ \hhline{=::====}
         Transition metal dichalcogenides \newline (Refs \cite{xu18},\,\cite{ma19},\,\cite{kan19},\,\cite{tiw21},\,\cite{ma22}) & Binary compound & $ < 100$\,K & \checkmark & \ding{55} \\ \hhline{-||----}
         Twisted and corrugated graphene \newline (Refs \cite{he22},\,\cite{ho21}) & Elemental            & \textless 200 K       & \checkmark          & \checkmark     \\ \hline
         Weyl semimetals \newline (Ref \cite{min23}) & Ternary compound     & Room temperature      & \ding{55}         & \ding{55} \\ \hline
         Dirac semimetals \newline (Ref \cite{min23})  & Ternary compound     & Room temperature      & \ding{55}         & \ding{55}    \\ \hline
         Topological insulators \newline (Ref \cite{he22}) & Binary compound      & \textless 200 K       & \ding{55}         & \ding{55}    \\ \hline
         Oxide heterostructure \newline (Ref \cite{les22}) & Ternary compounds    & \textless 30 K        & \checkmark          & \ding{55}    \\ \hline
         Bismuth \newline (this work) & Elemental            & Room temperature      & \checkmark          & \checkmark     \\ 
         \hhline{=:b:====}
    \end{tabular}
\end{table*}

\clearpage

\newpage

\section*{Methods}

\subsection*{Sample fabrication}
For transport characterization, thin films of bismuth with nominal thickness of $100$\,nm were deposited onto thermally oxidized $0.5$-mm-thick Si with a $500$-nm-thick SiOx layer (CrysTec GmbH, Germany) and mechanically flexible Kapton polymer foils of $25$-$\mu$m thickness (DuPont, USA). For THz THG and FHG characterization, the samples were grown on SiO$_2$ quartz glass substrates (CrysTec GmbH, Germany) and capped with a $2$-nm-thick Au layer to prevent oxidation. It is known that gold thin films do not contribute to THz high-harmonic generation. 
Some of the Bi($100$\,nm)/Au($2$\,nm) thin films were annealed in vacuum (below $10^{-3}$\,mbar) at $200^{\circ}$C for $1$\,h and $3$\,h to investigate the effect of microstructure on THz THG. Furthermore, to study scaling of THz THG efficiency with the number of bismuth interfaces we prepared [Bi($50$\,nm)/Au($4$\,nm)]$_2$ and [Bi($25$\,nm)/Au($2$\,nm)]$_4$ heterostructures.
To investigate the nonlinear Hall effect in Pt thin films, we prepared Hall crosses of $5$-nm-thick Pt on Si/SiOx($500$\,nm) substrates. To investigate THz SHG, we prepared extended array of arcs of $100$-nm-thick Bi thin films and $20$-nm-thick Au thin films on high purity silicon substrates. The arcs for THz studies were fabricated of 100-nm-thick Bi and 20-nm-thick Au thin films and had inner radius of $3\,\mu$m and width of $1.2\,\mu$m. To study scaling of the geometric nonlinear Hall effect in arc-shaped Hall bars, we lithographically patterned $100$-nm-thick Bi thin film to obtain a sample accommodating three arcs of different inner radius of $3,\,6,\,9\,\mu$m and width of $1.4\,\mu$m.
The deposition was done using RF magnetron sputtering at room temperature (base pressure: better than $10^{-7}$mbar; Ar sputter pressure: $10^{-3}$mbar; deposition rate: $0.3$\,nm/s). 
Hall cross devices were prepared using conventional optical lithography and lift-off process. The substrate was pre-baked at 120$^{\circ}$C for $5$\,min and surface cleaned in oxygen plasma for $3$\,min. TiPrime adhesion promoter (MicroChemicals GmbH, Germany) was spin coated at $3000$\,rpm and soft baked at $115^{\circ}$C for $3$\,min. Image reversal photoresist AZ5214e (MicroChemicals GmbH, Germany) was spin coated at $6000$\,rpm for $30$\,s and soft baked on a hot plate at $100^{\circ}$C for $90$\,s. The samples were exposed using direct laser writer (DWL66, Heidelberg Instruments, Germany), post-baked at $115^{\circ}$C for $90$\,s and developed in $1:4$ solution of AZ351b developer (MicroChemicals GmbH, Germany) in deionized (DI) water. After thin film deposition, extra material was lifted-off in acetone to reveal the device structure and rinsed with isopropanol and DI water.
For transport characterization, to prevent oxidation of bismuth surface layers, the devices were covered with GE-varnish (Oxford Instruments, UK).

\subsection*{X-ray diffraction}
XRD studies were carried out using a Rigaku SmartLab $3$\,kW with a parallel beam of {Cu}-$K_{\alpha}$ radiation (wavelength: $1.542$\,\AA). Lattice parameter refinement was performed with SmartLab Studio II. Bismuth crystallites lateral size was estimated from the width of four most intense X-ray reflections using the Scherrer equation.

\subsection*{Electron microscopy}
Scanning electron microscopy (SEM) imaging and electron backscatter diffraction (EBSD) were performed in a Zeiss NVision 40 SEM equipped with a field emission electron cathode and a Bruker EBSD system with an e-Flash HR+ detector.
The acceleration voltage was set to $30$\,kV, the beam current to about $10$\,nA using a $120\,\mu$m aperture. The EBSD detector resolution was set to $320\,\times\,240$ pixels and the exposure time to at least $6\,\times\,17$\,ms per frame. EBSD mapping was done as a rectangular grid of $1000\,\times\,750$ points with a step size of $100$\,nm.
The evaluation of the EBSD data was done with the Bruker EBSD software ESPRIT. 
To have high confidence on correct EBSD pattern indexation, the minimum number of indexed bands was set to 9 and the maximum mean angular deviation to $1.1^{\circ}$. These conditions resulted in indexation rates of $24$\% and $22$\% for the as-deposited and annealed samples, respectively. A manual review of the EBSD patterns revealed that mostly, the EBSD patterns are different between adjacent mapping points. Hence, the grain size is estimated to be about $100$\,nm or below, while the number of grains contributing to the mapping data is estimated to be of order $10^5$.

\subsection*{High-field magnetoresistance}
Pulsed-field magnetoresistance measurements reported in Fig. \ref{fig:fig1}(c) were performed at Dresden High Magnetic Field Laboratory, the large scale facility of the Helmholtz-Zentrum Dresden-Rossendorf. The measurement was carried out in the longitudinal geometry by 4-terminal method (current parallel to magnetic field). The sample was fed with AC current of $2$\,mA at a frequency of $13$\,kHz. The voltage was acquired by a digitizer at  sampling rate of $1$\,MHz. The recorded wave-forms were analyzed by a numerical lock-in software. Magnetic field was obtained by integrating the voltage induced in a calibrated pick-up coil located in the vicinity of the sample.

\subsection*{THz study}
The experiments were performed at the Helmholtz-Zentrum Dresden-Rossendorf. One part of the experiments (Figs. 1(d), 3(f)) was performed with the THz accelerator-based facility TELBE, and another part of the experiments (Figs. 1(e), 1(f), 4(a-f)) was performed with THz pulses generated by a Ti:sapphire femtosecond laser system using a tilted pulse front technique. Linearly polarized THz pulses with fundamental frequency of $0.3$\,THz and $0.5$\,THz were used as an input driving field. The input THz beam was focused on the samples at normal incidence, and the transmitted THz waves were detected behind the sample. During measurements, the temperature of the sample was controlled using cryostat (OptistatCF-V) in vacuum. After transmission through the sample, the THz waves containing the remainder of the fundamental field, as well as the generated harmonics, were sent through a set of bandpass filters with the transmission band centered at the generated harmonic frequency. The THz field of higher harmonics was measured directly in the time domain by free-space electro-optic sampling (FEOS) in a $2$\,mm thick ZnTe crystal.
For FEOS THz detection, we used synchronized optical probe pulses of $35\,$fs duration, central wavelength $800\,$nm. 
To measure the absolute value of the conversion efficiency of the THz field of higher harmonics, $100\,\mu$m GaP and no bandpass filters were used.

\subsection*{Magnetotransport and Sheet resistance}
The magnetotransport measurements reported in Fig. \ref{fig:fig2}(b) were carried out of an extended 100-nm-thick Bi thin film using the Zero-Offset Hall preset of a Tensormeter measurement device (Tensor Instruments, HZDR Innovation GmbH, Germany), which provides the absolute transverse resistance without any offset. Magnetic field was applied perpendicular to the film plane. Temperature control in the range of $270$\,K -- $320$\,K was achieved using cryostat in vacuum. The charge carriers density was calculated assuming only electron conductivity as
$$n \, =  \, \dfrac{1}{R_\text{H} t q},$$
where $R_\text{H}$ -- Hall resistance reported in Fig. \ref{fig:fig2}(b), $t = 100$\,nm -- film thickness and $q$ -- electron charge.
Temperature evolution of the sheet resistance (Fig. \ref{fig:fig2}(c)) was measured of an extended 100-nm-thick Bi thin film in the classical van der Pauw configuration by placing 4 electrical contacts at the edges of the sample. This measurement geometry provides sheet resistance with a geometry factor $= 1$.

\subsection*{Harmonic transport measurement}
Electrical transport harmonic measurements reported in the main text were performed using HF2LI Lock-in (Zurich Instruments, Switzerland). The sinusoidal current at the fundamental frequency of 787\,Hz was sourced using internal generator. Amplitude and phase of the measured longitudinal and transverse voltages at fundamental frequency and second harmonic were recorded. For each sourcing current value, the data was averaged over one minute integration time. Metal shielding around the sample was used to minimize the influence of external RF parasitic signals. 
For comparison, in Supplementary Information, we report measurements taken with two other devices including Tensormeter RTM2 (Tensor Instruments, HZDR Innovation GmbH, Germany) and lock-in amplifier SR860 (Stanford Research Systems, USA).

We note that the compound (\textit{i.e.}, including oscillator output and signal input) nonlinearity of our tested lock-in amplifiers (Stanford SR860 and Zurich HF2LI) poses a limit for the assessment of small harmonic distortions in the presence of a large fundamental signals. This is of particular relevance for the detection of the second harmonic longitudinal voltage. The SR860 and HF2LI are medium to high frequency lock-in amplifiers with a reported total harmonic distortion of $-80$\,dB (\textit{i.e.} 4 orders of magnitude). As a result, we see a noise floor in the dominant harmonic distortion orders (\textit{i.e.}, 2nd harmonic) that is approximately $10^{-4}$ of the first harmonic fundamental signal. This is evidenced even for the case of commercial surface mounted device (SMD) metal film resistors, where lock-ins show signal at the level of $20\,\mu$V at the second harmonic longitudinal voltage (Supplementary Fig. 13). As this noise can also scale with current, it can be mistaken for an actual harmonic distortion caused by the sample's physics. We have validated the harmonic distortion signals in our data through the additional measurements with a Tensormeter RTM2 (Tensor Instruments, HZDR Innovation GmbH, Germany). While targeted at lower frequency range of below $50$\,kHz, this device offers an improved compound nonlinearity of max. $5$\,ppm, which is confirmed by measuring apparent harmonic distortions from a highly linear reference made of commercial resistors. This test reassures that harmonic distortion signals reported to be in excess of $10^{-6}$ and measured with the Tensormeter RTM2 originate from the studied samples. In particular, we report current-voltage curves of the longitudinal voltage at the second harmonic. The second harmonic signal of the longitudinal voltage is at the level of $<100\,$nV for Bi thin films (Supplementary Fig. 14) and for the reference SMD resistor (Supplementary Fig. 13).

\section*{Data availability}
All data that support the plots within this paper and other findings  of this study are available from the corresponding authors upon reasonable request.

\section*{Acknowledgements}
We thank Conrad Schubert (HZDR) for support with thin film deposition, Dr. Larysa Baraban (HZDR) for providing access to the measurement equipment for harmonic analysis,  and Prof. Olav Hellwig (HZDR, TU Chemnitz) for providing access to the XRD tool. 
This research was carried out in part 
at the Ion Beam Center and ELBE large-scale facilities at the Helmholtz-Zentrum Dresden-Rossendorf e.V., member of the Helmholtz Association. We acknowledge the support of the Dresden High Magnetic Field Laboratory at the Helmholtz-Zentrum Dresden-Rossendorf e.V., member of the European Magnetic Field Laboratory (EMFL).
This work is financed in part via the German Research Foundation (DFG) under Grants No. MA 5144/22-1, MA 5144/24-1, MA 5144/33-1, and European Commission Community Research and Development Information Service (project REGO; ID: 951887). 

\bigskip

\section*{Author contributions}
C.O. and D.M. developed the project idea. P.M. prepared samples and performed magnetotransport characterization with the support from Y.Z., T.K. and D.M. S.K., I.I., A.P., A.A., G.C.P., T.O., and J.-C.D. performed THz studies. P.C. carried out electron microscopy measurements. F.G. performed XRD measurements with support from I.V. C.O. developed theoretical description of the transport effects in Bi thin films. Y.S. carried out high-field magnetoresistance measurements with the support from I.V. The manuscript was written by C.O., P.M. and D.M. with the contribution from S.K., T.K., I.I., A.P., P.C. Y.Z., I.V., Y.S., F.G, A.A., G.C.P., T.O., and J.-C.D.. 
	
\section*{Competing interests}
The authors declare no competing interests.

\end{document}


\title{Supplementary Information: \\ Tunable room temperature nonlinear Hall effect \\ from the surfaces of elementary bismuth thin films}

\author{Pavlo~Makushko}
\affiliation{Helmholtz-Zentrum Dresden-Rossendorf e.V., Institute of Ion Beam Physics and Materials Research, 01328 Dresden, Germany}

\author{Sergey~Kovalev}
\affiliation{Helmholtz-Zentrum Dresden-Rossendorf e.V., Institute of Radiation Physics, 01328 Dresden, Germany}

\author{Yevhen~Zabila}
\affiliation{Helmholtz-Zentrum Dresden-Rossendorf e.V., Institute of Ion Beam Physics and Materials Research, 01328 Dresden, Germany}
\affiliation{The H. Niewodniczanski Institute of Nuclear Physics,
Polish Academy of Sciences, 31–342 Krakow, Poland}

\author{Igor~Ilyakov}
\affiliation{Helmholtz-Zentrum Dresden-Rossendorf e.V., Institute of Radiation Physics, 01328 Dresden, Germany}

\author{Alexey~Ponomaryov}
\affiliation{Helmholtz-Zentrum Dresden-Rossendorf e.V., Institute of Radiation Physics, 01328 Dresden, Germany}

\author{Atiqa~Arshad}
\affiliation{Helmholtz-Zentrum Dresden-Rossendorf e.V., Institute of Radiation Physics, 01328 Dresden, Germany} 

\author{Gulloo~Lal~Prajapati} 
\affiliation{Helmholtz-Zentrum Dresden-Rossendorf e.V., Institute of Radiation Physics, 01328 Dresden, Germany} 

\author{Thales~V.~A.~G.~de~Oliveira}
\affiliation{Helmholtz-Zentrum Dresden-Rossendorf e.V., Institute of Radiation Physics, 01328 Dresden, Germany} 

\author{Jan-Christoph~Deinert}
\affiliation{Helmholtz-Zentrum Dresden-Rossendorf e.V., Institute of Radiation Physics, 01328 Dresden, Germany} 

\author{Paul~Chekhonin}
\affiliation{Helmholtz-Zentrum Dresden-Rossendorf e.V., Institute of Ion Beam Physics and Materials Research, 01328 Dresden, Germany}
\affiliation{Helmholtz-Zentrum Dresden-Rossendorf e.V., Institute of Resource Ecology, 01328 Dresden, Germany}

\author{Igor~Veremchuk}
\affiliation{Helmholtz-Zentrum Dresden-Rossendorf e.V., Institute of Ion Beam Physics and Materials Research, 01328 Dresden, Germany}

\author{Tobias~Kosub}
\affiliation{Helmholtz-Zentrum Dresden-Rossendorf e.V., Institute of Ion Beam Physics and Materials Research, 01328 Dresden, Germany}

\author{Yurii~Skourski}
\affiliation{Helmholtz-Zentrum Dresden-Rossendorf e.V., Dresden High Magnetic Field Laboratory (HLD-EMFL), 01328 Dresden, Germany}

\author{Fabian~Ganss}
\affiliation{Helmholtz-Zentrum Dresden-Rossendorf e.V., Institute of Ion Beam Physics and Materials Research, 01328 Dresden, Germany}

\author{Denys~Makarov}
\email{d.makarov@hzdr.de}
\affiliation{Helmholtz-Zentrum Dresden-Rossendorf e.V., Institute of Ion Beam Physics and Materials Research, 01328 Dresden, Germany}

\author{Carmine~Ortix}
\email{cortix@unisa.it}
\affiliation{Dipartimento di Fisica ``E. R. Caianiello", Universit\`a di Salerno, IT-84084 Fisciano (SA), Italy}

\maketitle

%
%
%
%

%

\maketitle	

\tableofcontents

\clearpage

\section{Supplementary structural characterization}

\subsection{X-ray diffraction data}

\begin{figure*}[h]
	\begin{center}
		\includegraphics[width=\linewidth]{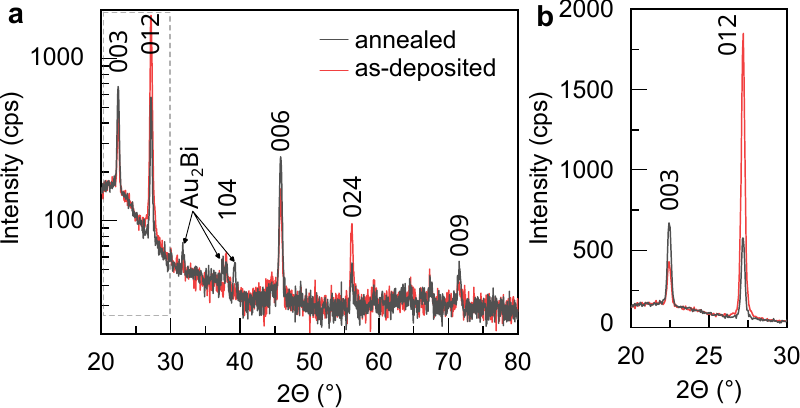}
	\end{center}
	\caption{\textbf{Structural characterization of $100$-nm-thick Bi thin films.} X-ray diffraction (XRD) patterns of the Bi($100$\,nm)/Au($2$\,nm) thin films. As-deposited thin film reveals high intensity of ($0\,0\,3$) and ($0\,0\,6$) reflections, indicating the presence of ($0\,0\,1$) texture. Vacuum annealing at $200^{\circ}$C enhances the ($0\,0\,1$) texture, in line with the increase of the ratio between ($0\,0\,3$) and ($0\,1\,2$) reflections. XRD data do not reveal the presence of the Au layer in the as-deposited sample due to low amount of material (only $2$\,nm). After annealing, low intensity peak of maldonite Au$_2$Bi appears, indicating thermally induced reaction of the gold capping layer with bismuth.}
	\label{fig:XRD}
\end{figure*}

\clearpage

\begin{table}[]
\caption{\textbf{Summary of the XRD data.} Lattice parameters, ratio of the integral intensity of XRD reflections and average grain size, estimated based on the Scherrer equation. Lattice parameters of bismuth in our thin films are close to ones for bulk material, indicating the absence of residual mechanical stresses in our thin films. Vacuum annealing enhances the ($0\,0\,l$) texture, retaining the microstructure of the samples (lattice parameters and grain size) unchanged.}
\label{tab:my-table}
\begin{tabular}{|l|l|l|l|l|l|}
\hline
\multicolumn{1}{|c|}{Sample} & $a$, \AA    & $c$, \AA      & $I_{003}$/$I_{012}$ & $I_{006}$/$I_{024}$ & Grain size, nm \\ \hline
PDF database                 & 4.54(7) & 11.86(2) & 6\%       & 86\%      & --             \\ \hline
as-deposited                 & 4.54    & 11.89    & 13\%      & 137\%     & 33             \\ \hline
$200^{\circ}$C $3$\,h                     & 4.54    & 11.87    & 93\%      & 972\%     & 34             \\ \hline
\end{tabular}
\end{table}

\clearpage

\subsection{Electron microscopy characterization}

\begin{figure*}[h]
	\begin{center}
		\includegraphics[width=1\linewidth]{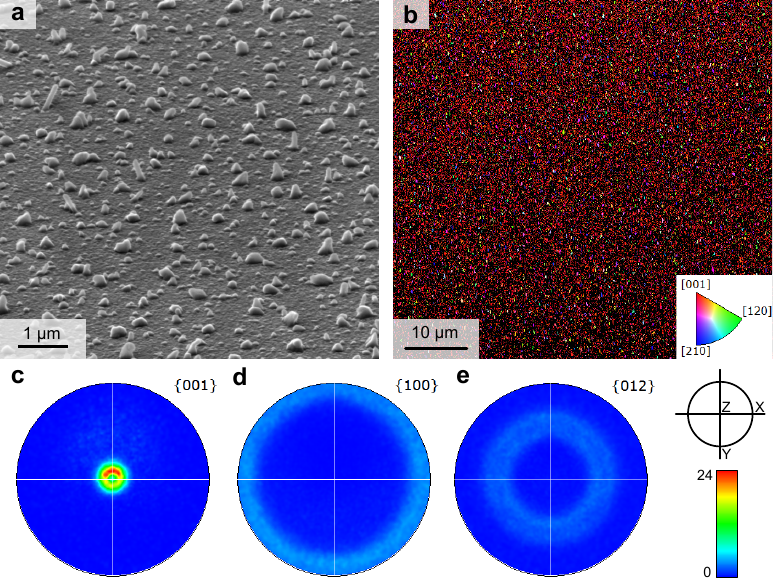}
	\end{center}
	\caption{\textbf{Microstructure characterization of the as-deposited $100$-nm-thick Bi thin film.} (a) High-resolution SEM secondary electron image of the sample surface (the sample is tilted by $70^{\circ}$). (b) Inverse pole figure for the thin film normal direction ($Z$), colors show crystallographic direction [$h\,k\,l$], which is parallel to the sample surface normal according to key figure. (c--e) Pole figures for selected crystallographic planes calculated from the EBSD data. The data indicate preferred ($0\,0\,1$) fiber texture in the thin film. The film is in-plane structurally isotropic. }
	\label{fig:EBSD1}
\end{figure*}

\clearpage

\begin{figure*}[h]
	\begin{center}
		\includegraphics[width=1\linewidth]{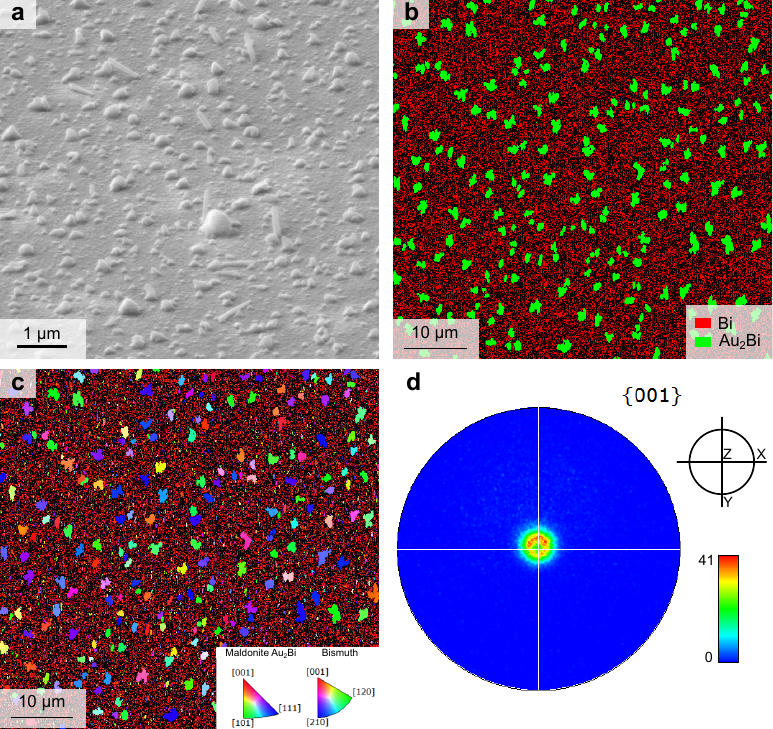}
	\end{center}
	\caption{\textbf{Microstructure characterization of the $100$-nm-thick Bi thin film after annealing at $200$$^{\circ}$C for $3$\,h.} (a) High-resolution SEM secondary electron image of the sample surface (the sample is tilted by $70^{\circ}$). (b) Phase map measured at the top surface of the thin film. Upon annealing, the $2$-nm-thick gold layer reacted with bismuth (red) and formed islands of maldonite Au$_2$Bi phase (green). (c) Inverse pole figure map for the thin film normal direction ($Z$) for both Bi and Au$_2$Bi phases. Colors show crystallographic direction [$h\,k\,l$], which is parallel to the sample surface normal according to the respective key figures. (d) Pole figure for the $\lbrace0\,0\,1\rbrace$ plane in bismuth calculated from the EBSD data. The figures indicate the presence of a ($0\,0\,1$) fiber texture in bismuth and random texture for maldonite grains.}
	\label{fig:EBSD2}
\end{figure*}

\clearpage

\section{Supplementary Theory Notes}

\subsection{Electronic properties and Berry curvature at the Bi ($1\,1\,1$) surface states}

Here, we derive the electronic properties of the ($1\,1\,1$) surface states of Bismuth using symmetry principles. 
We recall that in the remainder we use the rhombohedral notation for the crystal structure. 
Previous density functional theory calculations have shown that surface states in Bi are centered at the $\overline{\Gamma}$ point of the surface Brillouin zone. The minimal model Hamiltonian close to the surface effective spin-$\dfrac{1}{2}$ Kramers' doublet can be derived in a ${\bf k \cdot p}$ expansion accounting for all symmetry-allowed terms. To do so, we note that in the surface Kramers' doublet basis, the time-reversal symmetry can be represented as $\mathcal{T}=i\sigma_y \mathcal{K}$ with ${\mathcal K}$ the complex conjugation. The vertical mirror symmetry can be represented by $\mathcal{M}=i\sigma_x$. In the basis $\ket{\psi^{\uparrow\downarrow}}$, the threefold rotation operator takes the form $\mathcal{C}_3=e^{-i\sigma_z\pi/3}$. Under the operation of $\mathcal{C}_3$ and $\mathcal{M}$, momentum and spin transform as follows, 
\begin{align*}
&\mathcal{C}_3: & k_\pm &\rightarrow e^{\pm i 2\pi/3}k_\pm,& \sigma_\pm &\rightarrow e^{\pm i 2\pi/3}\sigma_\pm, & \sigma_z&\rightarrow\sigma_z\\
&\mathcal{M}: & k_+ &\rightarrow -k_-& \sigma_x &\rightarrow \sigma_x, & \sigma_{y,z}&\rightarrow-\sigma_{y,z}\\
\end{align*}
where $k_\pm=k_x\pm i k_y$ and $\sigma_\pm=\sigma_x\pm i \sigma_y$. The Hamiltonian must also be invariant under time reversal, which adds the constraint $\mathcal{H}(\mathbf{k})=\mathcal{T}\mathcal{H}(\mathbf{-k})\mathcal{T}^{-1}=\sigma_y\mathcal{H}^*(\mathbf{-k})\sigma_y$.

At linear order in the momentum $\mathbf{k}$, the minimal two-band Hamiltonian for a Kramers' related pair of bands corresponds to a Dirac cone:
\begin{equation}
\mathcal{H}_\textrm{TI}(\mathbf{k})= \hbar v_F \, \left(\sigma_x k_y - \sigma_y k_x \right) 
\label{eq:HTI}
\end{equation}
where $v_F$ is the Fermi velocity. 
The Hamiltonian above does not capture crystalline anisotropy effects. 
The first symmetry-allowed term accounting for crystalline anisotropy is third order in momentum and takes the form,
\begin{equation}
\mathcal{H}_w(\mathbf{k})=\frac{\lambda}{2}(k_+^3+k_-^3)\sigma_z \,.
\label{eq:Hwarping}
\end{equation}
This warping Hamiltonian is proportional to the Pauli matrix $\sigma_z$, which is crucial to obtain a non-zero Berry curvature and leads to out-of-plane spin textures. The Hamiltonian ${\mathcal H}=\mathcal{H}_\textrm{TI} + \mathcal{H}_w$ realizes the hexagonally warped Dirac cone~\cite{fu09} originally predicted to appear at the surface of the Bi$_2$Te$_3$. 
However, such an anomalous state is allowed only in strong three-dimensional topological insulators. Although recent reports~\cite{nay19} suggest that Bi can have a non-trivial ${\mathbb Z}_2$ time-reversal-protected topology, previous studies have indicated that this elemental material has a trivial strong topological index, and rather realizes a higher-order topological insulator~\cite{schind18}. 
The trivial value of the strong topological index implies that the effective surface Hamiltonian must be equipped with a regularizing term $\propto {\bf k}^2$ such that at each Fermi energy there is an even number of surface Kramers' pairs~\cite{KaneHasanrev}. As a result, the effective surface Hamiltonian reads 
\begin{equation}
\mathcal{H}(\mathbf{k})= \dfrac{\hbar^2 k^2}{2 m^{\star}} + \hbar v_F \, \left(\sigma_x k_y - \sigma_y k_x \right) + \frac{\lambda}{2}(k_+^3+k_-^3)\sigma_z, 
\label{eq:Hsurface}
\end{equation}
which corresponds to the Hamiltonian of a hexagonally warped Rashba two-dimensional electron system. 

The Berry curvature associated with this effective Hamiltonian can be calculated by rewriting Eq.~\ref{eq:Hsurface} as 
$\mathcal{H}(\mathbf{k})=\mathbf{d}(\mathbf{k})\cdot\boldsymbol{\sigma}+ 
 \mathbf{ \hbar^2 k}^2 \sigma_0 / (2 m)$.  
Here, ${\mathbf d}$ is a momentum dependent vector, which for our specific model has components ${\mathbf d}=\left\{-v_F k_y, v_F k_x, \lambda \left(k_{+}^3 + k_{-}^3 \right)/2 \right\}$.
In terms of the ${\mathbf d}$ vector, the Berry curvature can be expressed as $\Omega^\pm_z(\mathbf{k})=\pm\hat{\mathbf{d}}\cdot (\partial_{k_x}\hat{\mathbf{d}}\times\partial_{k_y}\hat{\mathbf{d} })/2$ with $\hat{\mathbf{d} } = \mathbf{d} / \left|\mathbf{d} \right|$. As a result, we have that for the surface effective Hamiltonian Eq.~\ref{eq:Hsurface} the Berry curvature is 
\begin{equation}
\Omega_\pm^{z}(k,\theta_{\bf k})=\pm\frac{2 \sqrt{2} \lambda  \hbar^2 v_F^2 k^3 \cos (3 \theta_{\bf k} )}{\left[2 \hbar^2 v_F^2 k^2+\lambda ^2 k^6 \cos (6 \theta_{\bf k} )+\lambda ^2 k^6\right]^{3/2}}\,, 
\label{eq:OmegaB}
\end{equation}
where $\theta$ is the polar angle in momentum space, and the $\pm$ sign distinguishes the two surface bands. 
The Berry curvature is well defined in each point except the origin where the bands are degenerate. 
 Note that the constraints set by time reversal symmetry and the three-fold rotational symmetry are satisfied as can be verified upon a closer inspection of Eq.~(\ref{eq:OmegaB}). Moreover, $\Omega_\pm^z(k,\theta_{\bf k})$ vanishes along the mirror lines thus proving that the Berry curvature is entirely driven by crystalline anisotropy terms. 
 The Berry curvature triple (BCT)~\cite{he21} is instead defined as a convolution of the Berry curvature with the angular form factor for the threefold rotation symmetry: 
 \begin{equation}
 \textrm{BCT}(\epsilon_F)=2 \pi \hbar \int \dfrac{d^2 {\bf k}}{(2 \pi)^2} \delta\left(\epsilon_F - \epsilon \right) \Omega_\pm^{z}(k,\theta_{\bf k}) \cos{\left(3 \theta_{\bf k} \right)} 
 \label{eq:BCTdef}
 \end{equation}
This quantity has dimension of time and, as shown below, governs the disorder-induced contribution to the nonlinear Hall effect.

\subsection{Second-order scattering rate and transport scattering time}
In this section, we derive the second-order scattering rate and the transport scattering time at the $(111)$ surfaces of Bi considering long-ranged impurities where the impurity potential corresponds to a screened Coulomb interaction.
To compute the scattering rates we 
write the normalized wavefunctions for the two surface state bands as 
\begin{eqnarray}
\ket{{\bf k}_{-}}&=& \left(\begin{array}{c} \sin{\dfrac{\varphi_{\bf k}}{2}} \\    i \cos{\dfrac{\varphi_{\bf k}}{2}} \exp{\left(i \theta_{\bf k}\right)} \end{array} \right) \exp{\left(i {\bf k \cdot r}\right)} \\  \ket{{\bf k}_{+}}& =& \left(\begin{array}{c} \cos{\dfrac{\varphi_{\bf k}}{2}} \\  - i \sin{\dfrac{\varphi_{\bf k}}{2}} \exp{\left(i \theta_{\bf k}\right)} \end{array} \right) \exp{\left(i {\bf k \cdot r}\right)},
\end{eqnarray}
where we introduced the angle $\varphi_{\bf k}$ defined by 
\begin{equation}
\cos{\varphi_{\bf k}}= \dfrac{\lambda  k^3 \, \cos{3 \theta_{\bf k}}}{\sqrt{\hbar^2 v_F^2 k^2 +\lambda^2 k^6 \cos^2{3 \theta_k}}}
\end{equation}

We next introduce the ``Dirac" energy $\epsilon_0({\bf k})=\sqrt{\hbar^2 v_F^2 k^2 +\lambda^2 k^6 \cos^2{3 \theta_k}}$ and the momentum-dependent Dirac mass term $m({\bf k})= \lambda k^3 \cos{3 \theta_k}$. In terms of these quantities, the generic Bloch state can be written as 
\begin{equation}
\ket{{\bf k}_{\mu}}= \dfrac{1}{\sqrt{2 \epsilon_0 ({\bf k})}} \left(\begin{array}{c} \sqrt{\epsilon_0({\bf k}) + \mu m({\bf k})} \\   -i \mu \sqrt{\epsilon_0({\bf k}) - \mu m({\bf k})} \exp{\left(i \theta_{\bf k}\right)} \end{array} \right) \exp{\left(i {\bf k \cdot r}\right)} 
\end{equation}
where $\mu=\pm 1$ distinguishes between the inner ($\mu=1$) and the outer ($\mu=-1$) warped Rashba bands. 
To make further progress, it is convenient to treat the warping term parametrized by $\lambda$ as a perturbation and hence expand in powers of $\lambda$. To linear order, we can then write the eigenstates as 
\begin{equation}
\ket{{\bf k}_{\mu}}= \dfrac{1}{\sqrt{2}}  \left(\begin{array}{c} 1 + \mu M({\bf k}) \\   -i \left( \mu - M({\bf k}) \right)  \exp{\left(i \theta_{\bf k}\right)} \end{array} \right) \exp{\left(i {\bf k \cdot r}\right)}, 
\end{equation}
where we introduced the dimensionless Dirac mass $M({\bf k})= \lambda k^2 \cos{3 \theta_k} / (2 \hbar v_F)$. 
The matrix element of the Coulomb impurity potential up to linear order in the Dirac mass is given by 
\begin{eqnarray}
\bra{{\bf k}_\mu} \mathcal V({\bf r}) \ket{{\bf k^{\prime}}_{\mu^{\prime}}} &=& \dfrac{V_0^{{\bf k}, {\bf k^{\prime}}}}{2} \left[ 1+ \mu \mu^{\prime} \mathrm{e}^{i (\theta_{\bf k^{\prime}} -\theta_{\bf k})} + \left( \mu M({\bf k}) + \mu^{\prime} M({\bf k^{\prime}}) \right) \right.  \\ & & \left.  \left(1 - \mu \mu^{\prime} \mathrm{e}^{i (\theta_{\bf k^{\prime}} -\theta_{\bf k})} \right)  \right] {\mathcal K}({\bf k}_{\mu}, {\bf k^{\prime}}_{\mu^{\prime}}) \nonumber
\end{eqnarray}
In the equation above, we have introduced $V_0^{{\bf k}, {\bf k^{\prime}}}= \sum_j 2 \pi Q e^2 / (4 \pi \epsilon_0 \epsilon) \, \mathrm{e}^{i ({\bf k} - {\bf k^{\prime}})\cdot {\bf r}_j}$, and the Coulomb kernel ${\mathcal K}({\bf k}_\mu, {\bf k^{\prime}}_{\mu^{\prime}})= \left(\left[k_\mu^2 + k^{\prime \,\,2}_{\mu^{\prime}} - 2 k_{\mu} k^{\prime}_{\mu^{\prime}} \cos{(\theta_{\bf k^{\prime}} -\theta_{\bf k})} \right]^{1/2} + q_{TF}\right)^{-1}$ where $q_{TF}$ is the Thomas-Fermi wavevector.

The symmetric part of the scattering rate is given by~\cite{sin07,du19,Ortix2021}
$$\omega^{(2)}_{l ; l^{\prime}} = \dfrac{2 \pi}{\hbar} \braket{|\bra{{\bf k}_\mu} \mathcal V({\bf r}) \ket{{\bf k^{\prime}}_{\mu^{\prime}}}|^2}_{dis} \, \, \delta( \epsilon_l - \epsilon_{l^{\prime}})$$
where $l=\left\{{\bf k}, \mu \right\}$. Let us introduce the dimensionless parameter $\alpha=Q e^2 / (4 \pi \epsilon_0 \epsilon \hbar v_F)$. At the zeroth order in the warping strength, the second-order scattering rate can be written as 
\begin{equation}
\omega^{(2)}_{l ; l^{\prime}}= 4 \pi^3 n_i \hbar v_F^2 \alpha^2 \left[1+ \mu \mu^{\prime} \cos{(\theta_{\bf k^{\prime}} -\theta_{\bf k})} \right]  \, {\mathcal K}({\bf k}_{\mu}, {\bf k^{\prime}}_{\mu^{\prime}})^2 \, \delta( \epsilon_l - \epsilon_{l^{\prime}})
\end{equation}
where $n_i$ is the impurity concentration.

Up to order $\lambda^2$, the Fermi surfaces can be considered circular with radius $k_{F \pm}$. 
Furthermore, the circular Fermi lines radii are related to each other by $k_{F_{\mu}}=k_{F_{\mu^{\prime}}} - 2 m v_F \mu / \hbar $. The momentum shift between the two parabolas proportional to the spin-orbit coupling strength defines the characteristic momentum $k_{SO}=2 m v_F / \hbar$. It is therefore convenient to introduce the shifted momenta ${\widetilde {\bf k}}={\bf k} + \mu k_{SO}/2$. Note that at the Fermi energy the shifted wavevector $\left| {\widetilde {\bf k}} \right|=\overline{k}_F = \sum_\mu k_{F_{\mu}} / 2$ is independent of the band index. 
The Coulomb kernel can be now written as a series expansion in $k_{SO}/ \overline{k}_F$. At the leading order we have 
\begin{equation}
{\mathcal K}^{-1}({\bf k}_{\mu}, {\bf k^{\prime}}_{\mu^{\prime}})= \overline{k}_F \left[2 \left(1-\cos{(\theta_{\bf k^{\prime}} -\theta_{\bf k})} \right) \right]^{1/2}  +q_{TF} 
\end{equation}
Consequently, the second-order scattering rate can be expressed as 
\begin{equation}
\omega^{(2)}_{l ; l^{\prime}} = \dfrac{4 \pi^3 n_i \hbar v_F^2 \alpha^2 \left[1 + \mu \mu^{\prime} \cos{(\theta_{\bf k^{\prime}} -\theta_{\bf k})} \right] }{\overline{k}_{F}^2 \left(2 \left|\sin{\dfrac{\theta_{\bf k}-\theta_{\bf k^{\prime}}}{2}}\right| + \dfrac{q_{TF}}{\overline{k}_{F}} \right)^2}   
\end{equation}

Then we determine the transport scattering time. 
We start by recalling that the transport scattering time can be found from the solution of the self-consistent equation 
\begin{equation}
e {\bf E} \cdot {{\bf v}_l} \dfrac{\partial f_l^0}{\partial \epsilon_l} = - \sum_{l^{\prime}} \omega^{(2)}_{l ; l^{\prime}} \left[-e {\bf E} \cdot {{\bf v}_l}~\tau^{tr}_{l} \dfrac{\partial f_l^0}{\partial \epsilon_l} + e {\bf E} \cdot {{\bf v}_{l^{\prime}}}~\tau^{tr}_{l^{\prime}} \dfrac{\partial f_{l^{\prime}}^0}{\partial \epsilon_{l^{\prime}}}  \right] 
\end{equation}
The equation above can be simplified by noticing that the second-order scattering rate conserves the energy and the equilibrium distribution function depends only on energy. Therefore, we are left with the set of self-consistent equations for the group velocities ${\bf v}_l=\partial \epsilon_l / (\hbar \partial {\bf k})$ 
\begin{equation}
{\bf v}_l= \tau_l^{tr} {\bf v}_l \sum_{l^{\prime}} \omega^{(2)}_{l ; l^{\prime}} - \sum_{l^{\prime}} \omega^{(2)}_{l ; l^{\prime}} {{\bf v}_{l^{\prime}}}~\tau^{tr}_{l^{\prime}}
\end{equation}
We follow Ref.~\cite{bro16} and define the quasiparticle scattering times $\dfrac{1}{\tau_l}=\sum_{l^{\prime}} \omega^{(2)}_{l ; l^{\prime}}$. 
Using this definition and the fact that upon momentum integration only the component of ${\bf v}_{l^{\prime}}$ parallel to ${\bf v}_l$ survives, we obtain that the equation above can be recast as 
\begin{equation}
\dfrac{\tau_l^{tr}}{\tau_l}=1+\sum_{l^{\prime}} \omega^{(2)}_{l ; l^{\prime}} ~\tau^{tr}_{l^{\prime}} \dfrac{{\bf v}_{l^{\prime}} \cdot \hat{v}_l}{|{\bf v}_l|}
\end{equation}
As noticed in Ref.~\cite{bro16}, this is a set of coupled equations for two band-dependent transport scattering times. 
At the zeroth order in $k_{SO}/\overline{k}_F$, the magnitude of the two band velocities are equal. Hence, we can derive a constant transport scattering time
\begin{equation}
\dfrac{1}{\tau^{tr}}=\sum_{l^{\prime}} \omega^{(2)}_{l ; l^{\prime}} \left(1- \cos{(\theta_{\bf k^{\prime}} -\theta_{\bf k})} \right)
\end{equation}
Substituting in the equation above the expression for the second-order scattering rate, the transport scattering time reads
\begin{eqnarray}
\dfrac{1}{\tau^{tr}}&=& \dfrac{2 \pi m n_i v_F^2 \alpha^2}{\hbar \overline{k}_{F}^2} c^{S}\left(\dfrac{q_{TF}}{\overline{k}_F} \right)
\end{eqnarray}
where the numerical factor 
\begin{equation}
c^S\left(\dfrac{q_{TF}}{\overline{k}_F} \right) = \int d \theta_{\bf k^{\prime}} \dfrac{\left(1- \cos{(\theta_{\bf k^{\prime}} -\theta_{\bf k})} \right)}{\left(2 \left|\sin{\dfrac{\theta_{\bf k}-\theta_{\bf k^{\prime}}}{2}}\right| + \dfrac{q_{TF}}{\overline{k}_{F}} \right)^2}
\end{equation}
and in the limit of unscreened impurities reduces to $c^S(q_{TF}/\overline{k}_F)=\pi$. In the opposite limit of $\delta$-like impurities, {\it i.e.} for $q_{TF} \gg \overline{k}_F$, the numerical factor $c^S(q_{TF}/\overline{k}_F) \simeq 2 \pi \overline{k}_F^2 / q_{TF}^2$, as expected.

\subsection{Third-order scattering rate and skew contribution to the nonlinear Hall effect}
In the following, we consider the third-order scattering rate (see for instance Ref.~\cite{sin07}) given by 
\begin{equation}
\omega^{(3)}_{l ; l^{\prime}}=\dfrac{2 \pi}{\hbar} \left( \sum_{l^{\prime \prime}} \dfrac{\braket{\mathcal{V}_{l l^{\prime}} \mathcal{V}_{l^{\prime} l^{\prime \prime}}\mathcal{V}_{l^{\prime \prime} l}}_{dis}}{\epsilon_l - \epsilon_{l^{\prime \prime}} - i \eta} + \mathrm{c.c.} \right) \delta(\epsilon_l - \epsilon_{l^{\prime}})
\label{eq:scattrate3}
\end{equation}
We note that the term in parenthesis can be simplified as 
\begin{equation}
\mathit{P}\left(\sum_{l^{\prime \prime}}  \dfrac{2 \mathrm{Re} \braket{\mathcal{V}_{l l^{\prime}} \mathcal{V}_{l^{\prime} l^{\prime \prime}}\mathcal{V}_{l^{\prime \prime} l}}_{dis}}{\epsilon_l - \epsilon_{l^{\prime \prime}}} \right) - 2 \pi \sum_{l^{\prime \prime}} \mathrm{Im} \braket{\mathcal{V}_{l l^{\prime}} \mathcal{V}_{l^{\prime} l^{\prime \prime}}\mathcal{V}_{l^{\prime \prime} l}}_{dis} \delta(\epsilon_l - \epsilon_{l^{\prime \prime}})
\end{equation}
The first term in the equation above does not contribute to the antisymmetric part of the third-order scattering rate. It is in fact symmetric under the $l \longleftrightarrow l^{\prime}$ once considering the Dirac $\delta$ function appearing in Eq.~\ref{eq:scattrate3}. As a result, we can write the antisymmetric third-order scattering term as 
\begin{equation}
\omega^{(3 a)}_{l ; l^{\prime}}=-\dfrac{(2 \pi)^2}{\hbar} \sum_{l^{\prime \prime}} \mathrm{Im} \braket{\mathcal{V}_{l l^{\prime}} \mathcal{V}_{l^{\prime} l^{\prime \prime}}\mathcal{V}_{l^{\prime \prime} l}}_{dis} \delta(\epsilon_l - \epsilon_{l^{\prime \prime}}) \delta(\epsilon_l - \epsilon_{l^{\prime}}) 
\end{equation}
After tedious but straightforward calculations, one finds that 
\begin{eqnarray}
\mathrm{Im} \braket{\mathcal{V}_{l l^{\prime}} \mathcal{V}_{l^{\prime} l^{\prime \prime}}\mathcal{V}_{l^{\prime \prime} l}}_{dis} & = & \dfrac{\left(2 \pi \hbar v_F \alpha\right)^3 n_i  \mu \mu^{\prime} \mu^{\prime \prime}}{2} \left[ \sin{\left(\theta_{q}-\theta_{k^{\prime}} \right)}  M({\bf k}_\mu) + \sin{\left(\theta_{k} - \theta_{q} \right)} M({\bf k^{\prime}}_{\mu^{\prime}})  \right. \nonumber \\ & & \left. -  \sin{\left(\theta_{k} - \theta_{k^{\prime}} \right)} M({\bf q}_{\mu^{\prime \prime}}) \right] {\mathcal K}({\bf k}_\mu, {\bf k^{\prime}}_{\mu^{\prime}}) {\mathcal K}({\bf k}_\mu, {\bf q}_{\mu^{\prime \prime}}) {\mathcal K}({\bf k^{\prime}}_{\mu^{\prime}}, {\bf q}_{\mu^{\prime \prime}}) 
\end{eqnarray}
Inserting this expression in the relation for the antisymmetric third-order scattering rate, we have 
\begin{eqnarray}
\omega^{\mu, \mu^{\prime} \,  (3 a)}_{{\bf k} ; {\bf k^{\prime}}}&=& -4 \pi^3 \hbar^2 \alpha^3 v_F^3 n_i \mu \mu^{\prime} \sum_{\mu^{\prime \prime}} \mu^{\prime \prime} \int q \, dq\, d\theta_{\bf q} \, \dfrac{\nu_{\mu^{\prime \prime}}(q)}{q} \delta(k_{F_{\mu^{\prime \prime}}} - q) \left[ \sin{\left(\theta_{q}-\theta_{k^{\prime}} \right)}  M({\bf k}_\mu)  \right. \nonumber \\ & & \left. + \sin{\left(\theta_{k} - \theta_{q} \right)} M({\bf k^{\prime}}_{\mu^{\prime}}) -  \sin{\left(\theta_{k} - \theta_{k^{\prime}} \right)} M({\bf q}_{\mu^{\prime \prime}}) \right] {\mathcal K}({\bf k}_{\mu}, {\bf k^{\prime}}_{\mu^{\prime}}) {\mathcal K}({\bf k}_{\mu}, {\bf q}_{\mu^{\prime \prime}}) {\mathcal K}({\bf k^{\prime}}_{\mu^{\prime}}, {\bf q}_{\mu^{\prime \prime}}) \nonumber \\ & &  \times \delta \left(\epsilon_{\bf k} - \epsilon_{\bf k^{\prime}} \right) 
\label{eq:scatterate32}
\end{eqnarray}
In the equation above, we have introduced the quantities related to the density of states 
$\nu_{\mu^{\prime \prime}}(q)^{-1}=\hbar^2 / m + \mu^{\prime \prime} \hbar v_F / q $. 
Note that the antisymmetric third-order scattering rate appears at linear order in the warping strength $\lambda$.

Let us first consider the limiting case of $\delta$-like impurities. In this case, the expression for the third-order scattering rate can be recast in the following form:
\begin{equation}
\omega_{{\bf k};{\bf k^{\prime}}}^{\mu, \mu^{\prime} \, (3a)} \propto \lambda \mu^{\prime} \sin{\left(\theta_{k} - \theta_{k^{\prime}} \right)}   \left[\nu_\mu k_{F_{\mu}}^2 - \nu_{\bar{\mu}} k_{F_{\bar{\mu}}}^2 \right] \int d \theta_q \cos{3 \theta_q} \times \delta \left(\epsilon_{\bf k} - \epsilon_{\bf k^{\prime}} \right) , 
\end{equation}
where $\bar{\mu}=-\mu$.
Importantly, the scattering rate vanishes because of the threefold symmetry of the dimensionless mass. This is in contrast with the case of a ferromagnetic Rashba model, where instead the absence of skew scattering~\cite{bor07} is due to the identity $\nu_{+}(k_{F_{+}}) / k_{F_{+}} \equiv \nu_{-}(k_{F_{-}})/ k_{F_{-}}$. Put differently, the vanishing of the skew scattering is not dependent on the number of occupied Kramers' pairs but is solely due to the threefold symmetry of the mass. 
A finite antisymmetric third-order scattering rate will therefore appear when instead of delta impurities we restore the screened Coulomb kernel for charged impurities.  

To explicitly show this, it is convenient to introduce, similarly to the second-order scattering rate, the shifted momentum ${\widetilde {\bf q}}={\bf q}+\mu k_{SO} /2$. 
For simplicity we calculate the Coulomb kernels for an equal amplitude $\overline{k}_F$. 
Then, Eq.~\ref{eq:scatterate32} can be recast as
\begin{eqnarray}
\omega_{{\bf k};{\bf k^{\prime}}}^{\mu, \mu^{\prime} \, (3a)} & \simeq & -4 \pi^3 \hbar^2 \alpha^3 v_F^3 n_i \mu \mu^{\prime} \sum_{\mu^{\prime \prime}} \mu^{\prime \prime}  \left({\overline k}_F - \dfrac{\mu^{\prime \prime} k_{SO}}{2} \right) \dfrac{m}{\hbar^2 \overline{k}_F} \, \dfrac{\lambda}{2 \hbar v_F} \dfrac{1}{\overline{k}_F^3} \int d \theta_{\bf q} \left[ \sin{\left(\theta_{{\bf q}}-\theta_{{\bf k}^{\prime}} \right)} \cos{3 \theta_{\bf k}}  \left(\overline{k}_F - \mu \dfrac{k_{SO}}{2}\right)^2  \right. \nonumber \\ & & \left. + \sin{\left(\theta_{{\bf k}} - \theta_{\bf q} \right)} \cos{3 \theta_{\bf k^{\prime}}} \left(\overline{k}_F - \mu^{\prime} \dfrac{k_{SO}}{2}\right)^2  -  \sin{\left(\theta_{{\bf k}} - \theta_{{\bf k}^{\prime}} \right)} \cos{3 \theta_{\bf q}} \left(\overline{k}_F - \mu^{\prime \prime} \dfrac{k_{SO}}{2}\right)^2 \right] \dfrac{1}{2 \left|\sin{\dfrac{\theta_{\bf k}-\theta_{\bf q}}{2}}\right| + \dfrac{q_{TF}}{\overline{k}_{F}}} \times \nonumber \\ & &  \dfrac{1}{2 \left|\sin{\dfrac{\theta_{\bf k^{\prime}}-\theta_{\bf q}}{2}}\right| + \dfrac{q_{TF}}{\overline{k}_{F}}} \, \, \, \dfrac{1}{2 \left|\sin{\dfrac{\theta_{\bf k}-\theta_{\bf k}^{\prime}}{2}}\right| + \dfrac{q_{TF}}{\overline{k}_{F}}} \times \delta \left(\epsilon_{\bf k} - \epsilon_{\bf k^{\prime}} \right) 
\end{eqnarray}
Here we have used the additional identity $\nu_{+}(k_{F_{+}}) / k_{F_{+}} \equiv m / (\hbar^2 \overline{k}_F)$.  
With this, we obtain that the non-vanishing antisymmetric third-order scattering rate can be conveniently written as
\begin{eqnarray}
\omega^{\mu, \mu^{\prime} \, (3 a)}_{{\bf k} ; {\bf k^{\prime}}}&=&\dfrac{2 \pi^3 \lambda  v_F^2 \alpha^3 n_i ~m  \mu \mu^{\prime}}{\hbar~\overline{k}_{F}}~\dfrac{\hat{z} \cdot ({\bf \hat{k}} \times {\bf \hat{k}^{\prime}})}{\left| {\bf \hat{k}} - {\bf \hat{k}^{\prime}}  \right|+\dfrac{q_{TF}}{\overline{k}_{F}}}  \delta \left(\epsilon_{\bf k} - \epsilon_{\bf k^{\prime}} \right) \left[\dfrac{k_{SO}}{\overline{k}_F} \, {\mathcal F}_{1}(\theta_{\bf k},\theta_{\bf k^{\prime}},\dfrac{q_{TF}}{\overline{k}_{F}}) + \dfrac{k_{SO}^3}{4 \overline{k}_F^3} {\mathcal F}_{3}(\theta_{\bf k},\theta_{\bf k^{\prime}},\dfrac{q_{TF}}{\overline{k}_{F}}) \right. \nonumber \\ & & \left. - \dfrac{k_{SO}^2}{\overline{k}_F^2} {\mathcal F}^{\mu \mu^{\prime}}_{2}(\theta_{\bf k},\theta_{\bf k^{\prime}},\dfrac{q_{TF}}{\overline{k}_{F}}) \right]
\label{eq:scaterate33}
\end{eqnarray}
where we have introduced three angular functions 
\begin{eqnarray}
{\mathcal F}_1(\theta_{\bf k},\theta_{\bf k^{\prime}},\dfrac{q_{TF}}{\overline{k}_{F}})&=&\int d \theta_{\bf q} \left[3 \cos{3 \theta_{\bf q}} - \dfrac{\sin{\left(\theta_{\bf q}- \theta_{\bf k^{\prime}}\right)} \cos{3 \theta_{\bf k}} + \sin{\left(\theta_{\bf k}- \theta_{\bf q}\right)} \cos{3 \theta_{\bf k^{\prime}}}}{\sin{\left(\theta_{\bf k}- \theta_{\bf k^{\prime}}\right)}}\right] \times \dfrac{1}{2 \left|\sin{\dfrac{\theta_{\bf k}-\theta_{\bf q}}{2}}\right| + \dfrac{q_{TF}}{\overline{k}_{F}}} \nonumber \\ & & \times \dfrac{1}{2 \left|\sin{\dfrac{\theta_{\bf k^{\prime}}-\theta_{\bf q}}{2}}\right| + \dfrac{q_{TF}}{\overline{k}_{F}}}
\end{eqnarray}
\begin{eqnarray}
{\mathcal F}_3(\theta_{\bf k},\theta_{\bf k^{\prime}},\dfrac{q_{TF}}{\overline{k}_{F}})&=&\int d \theta_{\bf q} \left[\cos{3 \theta_{\bf q}} - \dfrac{\sin{\left(\theta_{\bf q}- \theta_{\bf k^{\prime}}\right)} \cos{3 \theta_{\bf k}} + \sin{\left(\theta_{\bf k}- \theta_{\bf q}\right)} \cos{3 \theta_{\bf k^{\prime}}}}{\sin{\left(\theta_{\bf k}- \theta_{\bf k^{\prime}}\right)}}\right] \times \dfrac{1}{2 \left|\sin{\dfrac{\theta_{\bf k}-\theta_{\bf q}}{2}}\right| + \dfrac{q_{TF}}{\overline{k}_{F}}} \nonumber \\ & & \times \dfrac{1}{2 \left|\sin{\dfrac{\theta_{\bf k^{\prime}}-\theta_{\bf q}}{2}}\right| + \dfrac{q_{TF}}{\overline{k}_{F}}}
\end{eqnarray}
\begin{eqnarray}
{\mathcal F}_2^{\mu \mu^{\prime}}(\theta_{\bf k},\theta_{\bf k^{\prime}},\dfrac{q_{TF}}{\overline{k}_{F}})&=&\int d \theta_{\bf q} \dfrac{-\mu \sin{\left(\theta_{\bf q}- \theta_{\bf k^{\prime}}\right)} \cos{3 \theta_{\bf k}} -\mu^{\prime} \sin{\left(\theta_{\bf k}- \theta_{\bf q}\right)} \cos{3 \theta_{\bf k^{\prime}}}}{\sin{\left(\theta_{\bf k}- \theta_{\bf k^{\prime}}\right)}} \times \dfrac{1}{2 \left|\sin{\dfrac{\theta_{\bf k}-\theta_{\bf q}}{2}}\right| + \dfrac{q_{TF}}{\overline{k}_{F}}} \nonumber \\ & & \times \dfrac{1}{2 \left|\sin{\dfrac{\theta_{\bf k^{\prime}}-\theta_{\bf q}}{2}}\right| + \dfrac{q_{TF}}{\overline{k}_{F}}}
\end{eqnarray}

Next we follow the argument presented in Ref.~\cite{he21} and use that the largest contribution to the angular function, expanded in harmonic series, is given by ${\mathcal F}_{1,2,3}^{(\mu,\mu^{\prime})}(\theta_{\bf k},\theta_{\bf k^{\prime}},q_{TF}/\overline{k}_{F})= c_{1,2,3}^{A (\mu \mu^{\prime})}(q_{TF}/\overline{k}_{F}) \left(\cos{3 \theta_{\bf k}} + \cos{3 \theta_{\bf k^{\prime}}} \right)$. Numerical results for the $c^{A}$'s show that they decay for large $q_{TF}/\overline{k}_F$ is in agreement with our foregoing analysis for short-range impurities. Furthermore, we obtain that the constant $c^{A~1 -1}_{2}$ is numerically zero. Note that while the two terms governed by $c_{1,3}^A$ lead to equal intraband scattering rates opposite to the interband scattering rate, the two intraband scattering rates due to the $c_{2}^{A \pm1 \pm1}$ are opposite between each other.

Let us relate the antisymmetric scattering rate to the Berry curvature triple. Here we can use that in the linear order in the warping strength $\lambda$ the Berry curvature can be expressed as
\begin{equation}
\Omega^z_{\pm}\left(k, \theta_{\bf k} \right)= \pm \dfrac{\lambda \cos{3 \theta_{\bf k}}}{\hbar v_F} + \mathcal{O}\left(\lambda^2\right). 
\end{equation}
Inserting the expression above in the definition of the Berry curvature triple of Eq.~\ref{eq:BCTdef} for the two warped Rashba bands, we have 
\begin{equation}
BCT_{\mu}(\epsilon_F)= \dfrac{\mu \lambda m}{2 \hbar^2 v_F \overline{k}_F} \, \left(\overline{k}_F - \dfrac{\mu k_{SO}}{2}\right)
\end{equation}
The equation above also defines the average Berry curvature triple of our system given by 
\begin{equation}
\overline{BCT}(\epsilon_F)=-\dfrac{\lambda m k_{SO}}{4 \hbar^2 v_F k_F}
\end{equation}
and depends on the Fermi energy.
Substituting this expression in Eq.~\ref{eq:scaterate33}, we obtain the direct relation between the Berry curvature triple and the antisymmetric third-order scattering rate 
\begin{eqnarray}
\omega^{\mu, \mu^{\prime} \, (3 a)}_{{\bf k} ; {\bf k^{\prime}}}&=&- \dfrac{\left(2 \pi  \alpha v_F\right)^3~n_i~ 
\hbar~\overline{BCT}(\epsilon_F) \mu \mu^{\prime} }{k_{SO}}~\dfrac{\hat{z} \cdot ({\bf \hat{k}} \times {\bf \hat{k}^{\prime}})}{\left| {\bf \hat{k}} - {\bf \hat{k}^{\prime}}  \right|+\dfrac{q_{TF}}{\overline{k}_{F}}}  \, \left(\cos{3 \theta_{\bf k}} + \cos{3 \theta_{\bf k^{\prime}}} \right)  \delta \left(\epsilon_{\bf k} - \epsilon_{\bf k^{\prime}} \right) \times \nonumber \\ & & \left[\dfrac{k_{SO}}{\overline{k}_F} \, c_{1}^A(\dfrac{q_{TF}}{\overline{k}_{F}}) + \dfrac{k_{SO}^3}{4 \overline{k}_F^3} c_{3}^A(\dfrac{q_{TF}}{\overline{k}_{F}})  - \dfrac{k_{SO}^2}{\overline{k}_F^2} c^{A \mu \mu^{\prime}}_{2}(\dfrac{q_{TF}}{\overline{k}_{F}}) \right]
\label{eq:scaterate3final}
\end{eqnarray}

We can now proceed to evaluate the skew contribution to the nonlinear conductivity tensor $\chi_{\alpha \beta \gamma}$. As discussed below, the ${\mathcal C}_{3v}$ point group of the crystal implies that we need to compute a single component of the nonlinear conductivity tensor given by $\chi_{yxx}$. We therefore follow Ref.~\cite{du19,Ortix2021} and note that there exist two different terms in the skew scattering nonlinear conductivity $\chi_{yxx}$, the first of which is given by
\begin{equation}
\chi_{yxx}^{sk 1}= -\dfrac{e^3 \tau^3}{\hbar^3} \sum_{l l^{\prime}} \omega^{\mu, \mu^{\prime} \, (3 a)}_{{\bf k} ; {\bf k^{\prime}}} \left(\partial_{k_x} \partial_{k_y} \epsilon_l - \partial_{k^{\prime}_x} \partial_{k^{\prime}_y} \epsilon_{l^{\prime}} \right) \partial_{k_x} f_l^0,  
\end{equation}
where $f_l^0$ is the equilibrium Fermi-Dirac distribution function and $\tau$ is the transport scattering time evaluated before. At the linear order in the warping strength $\lambda$, it is sufficient to keep terms related to the energy of the Rashba bands up to $\lambda^0$. 
As a result, the first skew scattering contribution to the nonlinear Hall effect is given by
\begin{equation}
\chi_{yxx}^{sk 1}=- \dfrac{e^3 \tau^3~2~\left(2 \pi  \alpha \right)^3 v_F^4 m n_i \overline{BCT}(\epsilon_F)}{\hbar^3} \times \dfrac{k_{SO}}{\overline{k}_F^2} \times g^{sk 1} (\dfrac{q_{TF}}{\overline{k}_{F}})
\end{equation}
To obtain the equation above, we have used the symmetry properties of the various terms in the antisymmetric scattering rate and introduce a new constant given by the product of $c_2^{A 1 1}$ with the double integral over the angular dependence. We find that $g^{sk 1}$ is positive and of order one for $q_{TF}/\overline{k}_F \geq 0.4$.

Let us now compute the second skew scattering contribution to the nonlinear Hall effect. It is given by
\begin{equation}
\chi_{yxx}^{sk 2}= \dfrac{e^3 \tau^3}{\hbar^3} \sum_{l l^{\prime}} \omega^{\mu, \mu^{\prime} \, (3 a)}_{{\bf k} ; {\bf k^{\prime}}} \left( \partial_{k_y} \epsilon_l - \partial_{k^{\prime}_y}  \epsilon_{l^{\prime}} \right) \partial_{k_x} \partial_{k_x} f_l^0.  
\end{equation}
We can integrate by parts and thereby obtain that $\chi_{yxx}^{sk 2}=\chi_{yxx}^{sk 1}+\bar{\chi}_{yxx}^{sk 2}$ with 
\begin{equation}
\bar{\chi}_{yxx}^{sk 2}=-\dfrac{e^3 \tau^3}{\hbar^3} \sum_{l l^{\prime}}  \partial_{k_x} \omega^{\mu, \mu^{\prime} \, (3 a)}_{{\bf k} ; {\bf k^{\prime}}} \left( \partial_{k_y} \epsilon_l - \partial_{k^{\prime}_y}  \epsilon_{l^{\prime}} \right) \partial_{k_x} f_l^0
\end{equation}
We only explicitly consider the angular dependence of the scattering rate. Furthermore, using that the group velocity $\partial_{k_y} \epsilon_l = \sin{\theta_{\bf k}} \left(\hbar^2 k / m + \mu \hbar v_F\right)$ it can be found that only the component of the scattering rate $\propto c_2^{A \pm 1 \pm 1}$ gives a finite contribution at the leading order. With this, it follows that the total skew scattering contribution can be written as 
\begin{equation}
\chi_{yxx}^{sk}=- \dfrac{e^3 \tau^3~2~\left(2 \pi  \alpha \right)^3 v_F^4 m n_i \overline{BCT}(\epsilon_F)}{ \hbar^3} \times \dfrac{k_{SO}}{\overline{k}_F^2} \times \left(2 g^{sk 1} (\dfrac{q_{TF}}{\overline{k}_{F}}) + g^{sk 2} (\dfrac{q_{TF}}{\overline{k}_{F}}) \right)
\end{equation}
As for $g^{sk 1}$ also $g^{sk 2}$ is positive and of order one for $q_{TF}/\overline{k}_F \geq 0.4$.

As pointed out in Ref.~\cite{du19,Ortix2021}, the antisymmetric part of the fourth-order scattering rate provides an additional nonlinear skew scattering contribution to the nonlinear Hall effect. As we will show in Supplementary Theory Note E, also this term must be proportional to the warping strength $\lambda$ and thus to the Berry curvature triple.

\subsection{Coordinate shift and side jump contributions to the nonlinear Hall effect}
We first determine the coordinate shift whose gauge-invariant expression~\cite{sin07} is given by
\begin{equation}
\delta {\bf r}_{l l^{\prime}} = \braket{{\bf k}_\mu | i \partial_{\bf k} {\bf k}_\mu  } -  \braket{{\bf k}^{\prime}_{\mu^{\prime}} | i \partial_{{\bf k}^{\prime}} {\bf k}^{\prime}_{\mu^{\prime}}} - \left(\partial_{\bf k} + \partial_{{\bf k}^{\prime}} \right) \textrm{arg}\left[\braket{{\bf k}_\mu | {\bf k}_{\mu^{\prime}}}\right]
\end{equation}
To find the coordinate shift, we note that with the form of the wavefunction expanded up to linear order in the Dirac mass, the following relations hold
\begin{eqnarray}
\bra{{\bf k}_{\mu}} \partial_k \ket{{\bf k}_{\mu}} &=& M({\bf k}) \partial_{k} M({\bf k}) \nonumber \\
\bra{{\bf k}_{\mu}} \partial_{\theta_{\bf k}} \ket{{\bf k}_{\mu}}&=& M({\bf k}) \partial_{\theta_{\bf k}} M({\bf k}) + \dfrac{i}{2} \left[\mu - M({\bf k})\right]^2
\end{eqnarray}
In terms of these quantities, one can evaluate the coordinate shift contribution up to linear order in the warping strength $\lambda$
\begin{eqnarray}
i \bra{{\bf k}_{\mu}} \partial_{k_x} \ket{{\bf k}_{\mu}} &=& \dfrac{1}{2} \dfrac{\sin{\theta_{\bf k}}}{k} \left[1 - 2 \mu M({\bf k})\right] \nonumber \\
i \bra{{\bf k}_{\mu}} \partial_{k_y} \ket{{\bf k}_{\mu}} &=&  - \dfrac{1}{2} \dfrac{\cos{\theta_{\bf k}}}{k} \left[1 - 2 \mu M({\bf k})\right]
\end{eqnarray}
To evaluate the coordinate shift contribution $\propto \partial_{k_{x,y}} \textrm{arg}\left[\braket{{\bf k}_\mu | {\bf k^{\prime}_{\mu^{\prime}}}}  \right]$ we remember the following relation~\cite{du19}
\begin{equation}
\partial_x \textrm{arg}~A=\dfrac{i}{|A|^2} \left(\dfrac{1}{2} \partial_x |A|^2 - A^{\star} \partial_x A \right) 
\end{equation}

Moreover, we have that up to linear order in $\lambda$ 
\begin{eqnarray}
\braket{{\bf k}_\mu | {\bf k^{\prime}_{\mu^{\prime}}}}&=& \dfrac{1}{2} \left[\left(1+ \mu \mu^{\prime} \mathrm{e}^{i (\theta_{\bf k^{\prime}}- \theta_{\bf k})} \right) + M({\bf k}) \left(\mu - \mu^{\prime} \mathrm{e}^{i (\theta_{\bf k^{\prime}}- \theta_{\bf k})} \right) + M({\bf k^{\prime}}) \left(\mu^{\prime} - \mu \mathrm{e}^{i (\theta_{\bf k^{\prime}}- \theta_{\bf k})} \right)\right] \nonumber \\
\left|\braket{{\bf k}_\mu | {\bf k^{\prime}_{\mu^{\prime}}}}\right|^2&=& \dfrac{1}{2} \left(1+ \mu \mu^\prime \cos{(\theta_{\bf k^{\prime}} - \theta_{\bf k})} \right) 
\end{eqnarray}
\begin{eqnarray}
\partial_k \textrm{arg}\left[\braket{{\bf k}_\mu | {\bf k^{\prime}_{\mu^{\prime}}}}  \right]&=& - \dfrac{\partial_k M({\bf k})\mu^{\prime}}{\left(1 + \mu \mu^{\prime} \cos{(\theta_{\bf k^{\prime}} - \theta_{\bf k})} \right)} \, \sin{\left(\theta_{\bf k^{\prime}} - \theta_{\bf k} \right)}  \\ 
\partial_{k^{\prime}}\textrm{arg}\left[\braket{{\bf k}_\mu | {\bf k^{\prime}_{\mu^{\prime}}}}  \right]&=& - \dfrac{\partial_{k^{\prime}} M({\bf k}^{\prime})\mu}{\left(1 + \mu \mu^{\prime} \cos{(\theta_{\bf k^{\prime}} - \theta_{\bf k})} \right)} \, \sin{\left(\theta_{\bf k^{\prime}} - \theta_{\bf k} \right)} \\
\partial_{\theta_{\bf k}} \textrm{arg}\left[\braket{{\bf k}_\mu | {\bf k^{\prime}_{\mu^{\prime}}}}  \right]&=& -\dfrac{1}{2} - \dfrac{\partial_{\theta_k} M({\bf k})\mu^{\prime}}{\left(1 + \mu \mu^{\prime} \cos{(\theta_{\bf k^{\prime}} - \theta_{\bf k})} \right)} \, \sin{\left(\theta_{\bf k^{\prime}} - \theta_{\bf k} \right)} \nonumber \\ & & + \dfrac{\mu M({\bf k}) + \mu^{\prime} M({\bf k}^{\prime})}{\left(1 + \mu \mu^{\prime} \cos{(\theta_{\bf k^{\prime}} - \theta_{\bf k})} \right)} \\ 
\partial_{\theta_{{\bf k}^{\prime}}} \textrm{arg}\left[\braket{{\bf k}_\mu | {\bf k^{\prime}_{\mu^{\prime}}}}  \right]&=& \dfrac{1}{2} - \dfrac{\partial_{\theta_{k^{\prime}}} M({\bf k}^{\prime})\mu}{\left(1 + \mu \mu^{\prime} \cos{(\theta_{\bf k^{\prime}} - \theta_{\bf k})} \right)} \, \sin{\left(\theta_{\bf k^{\prime}} - \theta_{\bf k} \right)} \nonumber \\ & & - \dfrac{\mu M({\bf k}) + \mu^{\prime} M({\bf k}^{\prime})}{\left(1 + \mu \mu^{\prime} \cos{(\theta_{\bf k^{\prime}} - \theta_{\bf k})} \right)}
\end{eqnarray}

We will now show that the coordinate shift can be entirely expressed in terms of the Berry curvature triple just as the antisymmetric third-order scattering rate governing the skew scattering contribution to the nonlinear Hall effect. Let us first evaluate the contribution to the $\hat{x}$ component of the coordinate shift $\delta {\bf r}_{l l^{\prime}}^x$ given by $\braket{{\bf k}_\mu | i \partial_{k_x} {\bf k}_\mu  } - \partial_{k_x} \textrm{arg}\left[\braket{{\bf k}_\mu | {\bf k}_{\mu^{\prime}}}\right]$. Using that $\partial_k M({\bf k})=2 M({\bf k}) / k$ and trigonometric identities, we obtain that 
\begin{equation}
\braket{{\bf k}_\mu | i \partial_{k_x} {\bf k}_\mu  } - \partial_{k_x} \textrm{arg}\left[\braket{{\bf k}_\mu | {\bf k}_{\mu^{\prime}}}\right]=\dfrac{\mu^{\prime}}{2 k \left|\braket{{\bf k}_\mu | {\bf k^{\prime}_{\mu^{\prime}}}}\right|^2}\left[\sin{\theta_{{\bf k}}} M({\bf k}^\prime) - \sin{\theta_{{\bf k}^{\prime}}} M({\bf k}) + \dfrac{3 \lambda k^2}{2 \hbar v_F} \cos{2 \theta_{\bf k}} \sin{\left(\theta_{\bf k^{\prime}} - \theta_{\bf k} \right)} \right]
\end{equation}
We note that the last term in the equation above is derived from the angular dependence and the power low of the Dirac mass. Similarly, we have 
\begin{equation}
-  \braket{{\bf k}^{\prime}_{\mu^{\prime}} | i \partial_{k^\prime_x} {\bf k}^{\prime}_{\mu^{\prime}}} - \partial_{k_x^{\prime}}  \textrm{arg}\left[\braket{{\bf k}_\mu | {\bf k}_{\mu^{\prime}}}\right]=\dfrac{\mu}{2 k^{\prime} \left|\braket{{\bf k}_\mu | {\bf k^{\prime}_{\mu^{\prime}}}}\right|^2}\left[\sin{\theta_{{\bf k}}} M({\bf k}^\prime) - \sin{\theta_{{\bf k}^{\prime}}} M({\bf k}) + \dfrac{3 \lambda k^{\prime\,2}}{2 \hbar v_F} \cos{2 \theta_{{\bf k}^{\prime}}} \sin{\left(\theta_{\bf k^{\prime}} - \theta_{\bf k} \right)} \right]
\end{equation}
Combining these equations together, we have that 
\begin{eqnarray}
\delta {\bf r}_{l l^{\prime}}^x&=&-\dfrac{\hbar \overline{BCT}(\epsilon_F) \overline{k}_F}{m k_{SO} \left|\braket{{\bf k}_\mu | {\bf k^{\prime}_{\mu^{\prime}}}}\right|^2} \left[\mu k^{\prime} \left(\cos{3 \theta_{\bf k^{\prime}}} \sin{\theta_{\bf k}} + 3 \cos{2 \theta_{{\bf k}^{\prime}}} \sin{\left(\theta_{\bf k^{\prime}} - \theta_{\bf k} \right)} \right) - \mu \dfrac{k^2}{k^{\prime}} \cos{3 \theta_{\bf k}} \sin{\theta_{\bf k}^{\prime}} + \right. \nonumber  \\ & & \left. \mu^{\prime} k \left(-\cos{3 \theta_{\bf k} \sin{\theta_{\bf k}^{\prime}}} + 3 \cos{2 \theta_{{\bf k}}} \sin{\left(\theta_{\bf k^{\prime}} - \theta_{\bf k} \right)} \right)+ \mu^{\prime} \dfrac{k^{\prime 2}}{k} \cos{3 \theta_{\bf k}^{\prime}} \sin{\theta_{\bf k}}  \right]
\end{eqnarray}
A similar equation can be derived also for the $\hat{y}$ component of the coordinate shift. It has the same structure and most importantly is governed by the Berry curvature triple. One can then compute the side-jump velocity ${\bf v}_{sj}= \sum_{l^{\prime}} \omega_{l l^{\prime}}^{(2)} \delta~{\bf r}_{l^{\prime} l}$ and the two side-jump contributions to the nonlinear Hall effect whose expressions are given in Ref.~\cite{du19,Ortix2021}. We also note that there is an additional side-jump contribution related to the interband Berry connection that has no counterpart in linear response~\cite{xia19}. As we will show in Supplementary Theory Note E, also this term must be proportional to the warping strength $\lambda$ and thus to the Berry curvature triple. 

\subsection{Symmetry constraints on the nonlinear conductivity tensor}

\begin{figure*}[h!]
	\begin{center}
		\includegraphics[width=0.95\linewidth]{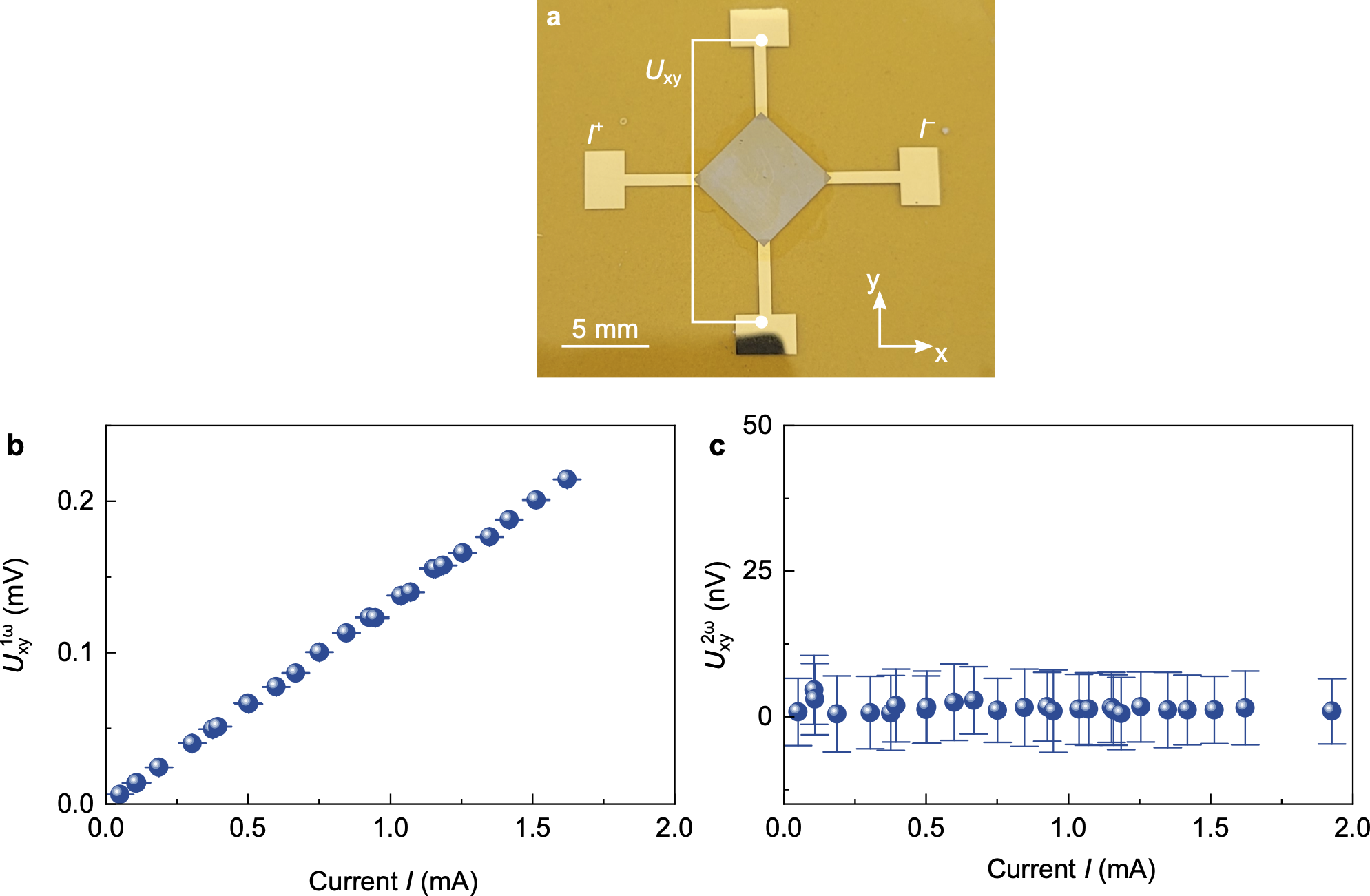}
	\end{center}
	\caption{\textbf{Harmonic measurement of a \textit{macroscopic} $100$-nm-thick Bi Hall cross prepared on a polymeric foil.} 
 (a) Photo of the macroscopic $100$-nm-thick Bi thin film structured into a Hall cross on mechanically flexible $25$-$\mu$m-thick Kapton foil (DuPont, USA). Lateral dimensions of the Hall cross are $5\times5$\,mm$^2$. The middle part of the device is covered with GE-varnish (Oxford Instruments, UK) to prevent bismuth thin film from oxidation. 
 Electrical measurements are done with Tensormeter RTM2 (Tensor Instruments, HZDR Innovation GmbH, Germany). Transverse voltage measured at the (b) first harmonic and (c) second harmonic. The measurements are done at the fundamental frequency of $787$\,Hz. Contact pads are realized with $100$-nm-thick gold thin films.
 }
	\label{fig:MacroscopicCross}
\end{figure*}

The nonlinear conductivity tensor is defined by the relation
\begin{equation} 
j_{\alpha}=\chi_{\alpha \beta \gamma} E_{\beta} \, E_{\gamma}
\end{equation}
We can derive the transformation rule of the nonlinear conductivity tensor under a generic point group symmetry represented by an orthogonal matrix 
by noticing that both the current $j$ and the driving electric field $E$ transform as vectors under a generic coordinate change. 
The transformation rule of the nonlinear conductivity tensor imply that in the presence of a ${\mathcal M}_x$  mirror symmetry, we have $\chi_{x x x}=\chi_{x y y }=\chi_{y x y}=\chi_{y y x}=0$. 
The additional threefold rotation symmetry in the ${\mathcal C}_{3v}$ point group symmetry implies that the non-zero component of the nonlinear conductivity tensor satisfies the relation $\chi_{x x y} = \chi_{x y x} = \chi_{y x x} = - \chi_{y y y}$. As a result, the nonlinear conductivity tensor is governed by a single quantity.
Note that for a driving electric field $E_{x}$ that is orthogonal to the vertical mirror, the generated nonlinear current is purely transverse. For a driving electric field at an angle $\phi$ with respect to ${\hat x}$, the generated nonlinear transverse current $j_T= {\bar \chi} E^2 \cos{3 \phi}$. The net transverse current in a polycrystalline sample is thus dependent on the distribution of the crystallites in-plane orientation. 
Note that the small statistics of grains in Hall crosses of finite area (about $1 \times 1$\,$\mu$m$^2$ cross area; grain size of about 30\,nm) guarantees the non-vanishing of the nonlinear transverse current independent of the distribution. 
This assumption is confirmed by studying macroscopic, \textit{i.e.} $5\times5$\,mm$^2$, Bi Hall cross (Supplementary Fig. \ref{fig:MacroscopicCross}(a)). In particular, transport data demonstrate that this sample does not reveal nonlinear transverse response (Supplementary Fig. \ref{fig:MacroscopicCross}(b,c)). 

Symmetry constraints on the nonlinear Hall effect can be also used to show the one-to-one correspondence between the Berry curvature of the Rashba bands and the disorder-mediated contributions introduced in the preceding Supplementary Theory Notes.
Consider the states at a surface containing an evenfold rotation symmetry, {\it i.e.} with point group ${\mathcal C}_{2v}$, ${\mathcal C}_{4v}$ or ${\mathcal C}_{6v}$. The effective Hamiltonian for the surface states up to quadratic order will correspond to a Rashba two-dimensional electron system. However, the trigonal warping term appearing at the $(111)$ surfaces is symmetry-forbidden and the Berry curvature is vanishing. And indeed the concomitant presence of the twofold rotation symmetry ${\mathcal C}_2$ and time-reversal symmetry ${\Theta}$ implies the presence of the additional combined antiunitary symmetry ${\mathcal C}_2 \Theta$ that squares to 1 and transforms ${\bf k}$ in itself.  
Since ${\mathcal C}_2 \Theta$ can be represented by a complex conjugation, we obtain that its presence forces the Berry curvature to vanish at any momentum. On the other hand, from the relation $j_{\alpha}= \chi_{\alpha \beta \gamma} E_{\beta} E_{\gamma} $, we have that the presence of a twofold rotation symmetry implies vanishing of all components of the nonlinear conductivity tensor in a two-dimensional system. Combining these results, we obtain that surface states realizing a Rashba two-dimensional electron system must have a vanishing nonlinear Hall effect in the absence of Berry curvature. Consequently, all disorder-mediated contributions to the nonlinear Hall must be directly related to the strength of the Berry curvature effect and, in particular, to the Berry curvature triple.

\clearpage

\subsection{On the difference between dissipative and dissipationless (non)linear transverse electrical responses.}

To elucidate the difference between dissipative and dissipationless transverse electrical responses, we first consider the linear transport regime. In particular let us assume that in a ${\mathcal C}_2$-symmetric material
a current is injected in a direction different from a principal crystallographic direction and a linear transverse voltage is obtained. The corresponding transverse conductivity  is related to the longitudinal conductivity by a rotation of the conductivity tensor. Hence it will be dissipative. 
Such dissipative transverse voltages can be also completely unrelated to longitudinal voltages.
Consider for example a crystalline system with a trivial ${\mathcal C}_1$ point group symmetry -- the only symmetry in the space group are discrete translations. 
This situation is realized in, for instance, $(111)$-grown LaAlO$_3$/SrTiO$_3$ interfaces [see Ref.~\cite{les22}] since the cubic-to-tetragonal structural phase transition of SrTiO$_3$ involving antiferrodistortive oxygen octahedron rotations breaks all mirror and rotation symmetries. The linear conductivity tensor is characterized by the presence of a transverse component, which in a constant relaxation time approximation, is related to the (non-vanishing after integration) velocity product $v_{x} v_{y}$. As a result, a transverse voltage can be produced even when a current is injected along a principal crystallographic direction. 

There is a key physical difference between conventional Hall voltages and these time-reversal symmetric transverse voltages. Transverse voltages can be in fact differentiated by looking at  whether or not they contribute to the dissipated power ${\bf j} \cdot {\bf E}$. An antisymmetric part of the transverse conductivity with $\sigma_{xy}=-\sigma_{yx}$ does not contribute to the dissipative power. 
On the contrary, the symmetric part of the transverse conductivity with $\sigma_{xy}=\sigma_{yx}$ is dissipative. These properties combined with Onsager relations $\sigma_{xy}(B)=\sigma_{yx}(-B)$ imply that any transverse voltages observed in time-reversal symmetric conditions must be dissipative. 

It would be then natural to associate Hall voltages with dissipationless transverse currents. This, however, is not the case. Consider the well-known planar Hall effect conventionally observed in ferromagnets with strong spin-orbit coupling. The planar Hall effect has gained significant attention in recent years in relation with time-reversal symmetric Weyl semimetals where it originates from the so-called chiral anomaly [see Ref.~\cite{nan17}]. In the planar Hall effect, coplanar electric and magnetic fields lead to a transverse voltage. However, such voltage is dissipative for the simple reason that it is unchanged if the direction of the applied magnetic field is reversed. Furthermore, the transverse voltage is connected to the anisotropic longitudinal magnetoresistance. 
This effect thus represents a prime example of a transverse electrical response with the same underlying mechanism as the longitudinal one and to which we conventionally refer to as Hall.

The question that now arises is whether the transverse nonlinear response, which we observe in our Bi thin films, can be dubbed as a nonlinear Hall voltage or should be called in more broad terms as electrical second-harmonic generation. The nonlinear transverse response of a threefold rotation symmetric system is dissipative due to the equivalence $\chi_{xyx}=-\chi_{yxx}$ [see Ref.~\cite{Ortix2021}] that prohibits a Berry curvature dipole. However, as shown above, it can be entirely related to another Berry curvature related quantity: the Berry curvature triple. 
Furthermore, the recent study Ref~\cite{Du2021} has dubbed the nonlinear electrical transverse responses appearing at the threefold symmetric surfaces of Bi$_2$Se$_3$  as nonlinear Hall. 
Finally, the classical geometry-induced second-order transverse responses has been also named Hall [see Ref.~\cite{sch19,gen22}] due to the bent electron trajectories similar to  those appearing in the conventional Hall effect. For these reasons, we refer to the nonlinear transverse response in Bi as nonlinear Hall.

\clearpage

\subsection{Scaling of the nonlinear Hall effect in Bi thin films}

In this section we investigate the scaling of the nonlinear Hall effect in polycrystalline bismuth thin films. 
For good metals, dynamic (electron-phonon) scattering sources can compete with the impurity scattering sources, analyzed in Supplementary Theory Notes B,C,D. In this respect, a scaling analysis of the nonlinear Hall effect can help in unveiling the relative role of static impurity and phonon scattering processes. 

\begin{figure*}[tbp]
	\begin{center}
\includegraphics[width=0.85\linewidth]{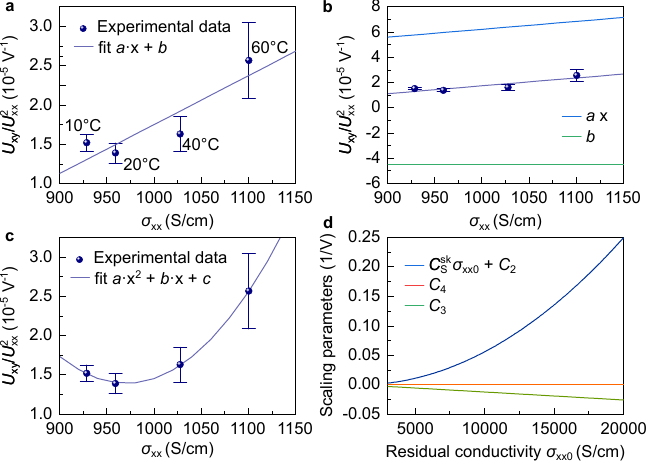} \end{center}
	\caption{\textbf{Scaling of the nonlinear transverse voltage $U_\text{xy}^{2\omega}$ vs. longitudinal \textit{bulk} conductivity $\sigma_\text{xx}$ of $100$-nm-thick bismuth thin films prepared on Si/SiOx($500$\,nm).} 
Nonlinear transverse voltage and longitudinal bulk conductivity are measured of the same Bi Hall cross structure upon varying the measurement temperature. Temperature is indicated in the figure panel for each data point. 
 In panel (a) the experimental data points are overlayed with the fit $U_\text{xy}/U_\text{xx}^2 = a\cdot\sigma_\text{xx}+b$. 
 Fit coefficients are: 
 $a = (6.2\pm2.2)\times10^{-8}\,\frac{\text{cm}}{\text{A}}$;
 $b = (4.5\pm2.3)\times10^{-5}\,\frac{1}{\text{V}}$.
 Panel (b) shows the individual components of the fit.
 In panel (c) the experimental data points are overlayed with the fit $U_\text{xy}/U_\text{xx}^2 = a\cdot\sigma_\text{xx}^2+b\cdot\sigma_\text{xx}+c$. Fit coefficients are: 
 $a = (6.9\pm0.3)\times10^{-10}\,\frac{\text{V}\cdot\text{cm}^2}{\text{A}^2}$;
 $b = (-1.35\pm0.05)\times10^{-6}\,\frac{\text{cm}}{\text{A}}$;
 $c = (6.7\pm0.3)\times10^{-4}\,\frac{1}{\text{V}}$.
 (d) Scaling terms for the static ($C_S^{sk} \sigma_{xx0} + C_2$), dynamic ($C_4$) and mixed ($C_3$) charge carriers scattering obtained from the fit shown in panel (c) vs residual conductivity $\sigma_\text{xx0}$.
 Transverse voltage $U_\text{xy}^{2\omega}$ is measured at the second harmonic. Longitudinal voltage $U_\text{xx}$ and longitudinal conductivity $\sigma_\text{xx}$ are measured at the first harmonic. Transport measurements are done at the fundamental frequency of $787$\,Hz. The measurement was done using lock-in amplifier HF2LI (Zurich Instruments AG, Switzerland). 
} 
	\label{fig:Scaling}
\end{figure*}

We follow Ref.~\cite{du19} and recall that for our system the absence of a Berry curvature dipole implies the existence of three different scaling terms in the nonlinear Hall conductivity, $\chi_{yxx}$. Nonlinear side jump contributes as $\chi_{yxx}^{sj} \propto \rho_i / \rho_{xx}^2$. Here $\rho_i$ is the resistivity of source $i$ [see Ref.~\cite{hou15}] that is connected to the total resistivity by Mathiessen's rule $\sum_i \rho_i=\rho_{xx}$. Note also that the nonlinear Hall contribution related to the interband Berry connection introduced in  Ref.~\cite{xia19} scales in the same manner as the side jump and can be thus included in $\chi_{yxx}^{sj}$. The third-order skew scattering (see Supplementary Theory Note D) instead scales as $\chi_{yxx}^{sk} \propto \rho_i / \rho_{xx}^3$, where the cubic dependence on $\rho_{xx}$ follows from the fact that $\chi_{yxx}^{sk} \propto \tau^3$ while the $\rho_i$ factor is associated to the third-order scattering rate. Finally, the skew scattering contribution due to the antisymmetric part of the fourth-order scattering rate scales as $\chi_{yxx}^{sk, 4} \propto \rho_i \rho_j / \rho_{xx}^3$. Using the relations $\sigma_{yy} E_y=\sigma_{xx} E_y=\chi_{yxx} E_x^2$, we can express the ratio between the second-harmonic transverse voltage $U_{xy}^{2 \omega}$ and the square of the first-harmonic longitudinal voltage $U_{xx}^{1 \omega}$ as
\begin{equation}
\dfrac{U_{xy}^{2 \omega}}{(U_{xx}^{1 \omega})^2}=\chi_{yxx} \rho_{xx}=\sum_i C_{i}^{sj} \dfrac{\rho_i}{\rho_{xx}} +  C_S^{sk} \dfrac{\rho_{S}}{\rho_{xx}^2} + \sum_{i j} C_{i j}^{sk,4} \dfrac{\rho_i \rho_j}{\rho_{xx}^2}
\label{eq:genscaling}
\end{equation}
This analysis is in agreement with Ref.~\cite{du19}. In the equation above, we introduced the coefficients $C$ quantifying the source-dependent side jump and skew scattering contributions. Moreover, we assumed that in the third-order skew scattering only the contribution due to static impurities $C_S$ is relevant [see Ref.~\cite{tia09,hou15}].

First, we assume that there is a single source of scattering due to static impurities. A similar assumption was made for Bi$_2$Se$_3$ in Ref.~\cite{he21}. In this case, Eq.~\ref{eq:genscaling} reduces to
\begin{equation}
\dfrac{U_{xy}^{2 \omega}}{(U_{xx}^{1 \omega})^2}=(C_S^{sj}+ C_S^{sk,4})+C_S^{sk} \sigma_{xx},
\end{equation}
where we have used that $\rho_S=\rho_{xx}$. 
We have fitted the experimental data obtained of a Bi Hall cross by measuring transport at different temperatures with the linear behavior of the equation above [see Supplementary Fig.~\ref{fig:Scaling}(a,b)]. We find that the skew scattering $C_{S}^{sk}$ and the side jump $C_{S}^{sj}$ coefficients have opposite sign and are comparable in magnitude. Despite the similar values of the longitudinal conductivities, there is thus a qualitative difference with the case of Bi$_2$Se$_3$ where the third-order skew scattering was argued to be more relevant than the side jump contribution and the fourth-order skew scattering. 
Importantly, the authors of Ref.\cite{he21} have considered a different type of scaling for the third-order skew scattering $\propto \sigma_{xx}^2$, in disagreement with our analysis and Ref.~\cite{du19}. 

We now show that such quadratic term $\propto \sigma_{xx}^2$ appears when considering two sources of scattering: static impurities and phonons. The resistivities associated to these two scattering sources correspond respectively to the zero-temperature residual resistivity $\rho_{xx0}$ and $\rho_{xxT}=\rho_{xx}-\rho_{xx0}$. Then the scaling of Eq.~\ref{eq:genscaling} can be written as 
\begin{equation}
\dfrac{U_{xy}^{2 \omega}}{(U_{xx}^{1 \omega})^2}=\dfrac{C_S^{sk} \sigma_{xx0} + C_2 + C_4 - C_3}{\sigma_{xx0}^2} \sigma_{xx}^2 + \dfrac{C_3 - 2 C_4}{\sigma_{xx0}} \sigma_{xx} + C_4 
\label{eq:scalingimpph}
\end{equation}
In the equation above, we introduced new scattering coefficients $C_{2,3,4}$ with 
\begin{eqnarray*}
C_2 &=& C_S^{sj} + C_{S S}^{sk, 4} \\ 
C_3 &=& C_S^{sj} + C_{ph}^{sj}+ C_{S ph}^{sk, 4} \\
C_4 &=& C_{ph}^{sj} + C_{ph ph}^{sk, 4}, 
\end{eqnarray*}
where the subscript $S,ph$ stands for static impurities and phonons, respectively.
Note that the coefficient $C_2$ depends only on static impurity scattering, similarly to $C_S^{sk}$. On the contrary, the coefficient $C_4$ depends only on dynamic scattering processes. Finally, $C_3$ is a mixed term. 
Taking into account the scaling of Eq.~\ref{eq:scalingimpph}, we obtain a much better fit, as reported in Supplementary Fig.~\ref{fig:Scaling}(c,d). Furthermore, from our fits we can estimate the dynamic parameter $C_4$, a purely static scattering parameter $C_S^{sk} \sigma_{xx0} + C_2$ as well as the mixed term $C_3$. In Supplementary Fig.~\ref{fig:Scaling}(e) we show their behavior as a function of the residual conductivity of bismuth, $\sigma_{xx0}$. For residual conductivities that are one order of magnitude larger than the measured room-temperature conductivity, the static scattering term $C_S^{sk} \sigma_{xx0} + C_2$ is much larger than the two other coefficients. This implies that for large enough residual conductivities the nonlinear Hall signal will be dominated by static impurity scattering. Note that in the literature values as high as $\sigma_{xx0} \simeq 10^6 $S$/$cm have been reported for the residual conductivity of Bi [see Ref.~\cite{uhe77}].

In conclusion, both our scaling analyses point to a dominant role played by static impurities with no particular indication of a preferred skew scattering mechanism. As discussed above, both side jumps and skew scattering mechanisms originate from the Berry curvature triple.
We also note that according to our scaling laws, which agree with Ref.~\cite{du19}, it would not be possible to individuate a predominant skew scattering mechanism in Bi$_2$Se$_3$ due to the quadratic scaling of $U_{xy}^{2 \omega}/(U_{xx}^{1 \omega})^2$.

\clearpage

\subsection{Thermal effects in bismuth thin films due to charge current}

To estimate relevance of thermal effects in our 100-nm-thick Bi films induced by charge current, we use a bimaterial thermal model proposed in \cite{Aviles2003}. The model considers a conductive thin film material through which the current flows and an electrically insulating substrate. The thermal state of the system is found by solving the heat balance equations in which the thermal energy supplied to the system by Joule resistive heating (i) increases its temperature and (ii) is lost via thermal conduction and convection to the environment.

Joule heating of an electrically resistive material with resistance $R$ due to flowing charge current $I$ is described with the following equation:
\begin{equation}
    \label{eqn:Therm1}
Q = R I^2 \Delta t = \rho \frac{l_f}{d_f a_f} I^2 \Delta t,
\end{equation}
where $\rho$ - resistivity of the thin film material, $\Delta t$ - time, and $l_f$, $a_f$ and $d_f$ stand for length, width and thickness of the conductive thin film. The energy $Q$ is accumulated by the material as thermal energy resulting in an increase of its temperature $T$ from the initial temperature $T_0$ for $\Delta T$:
\begin{equation}
    \label{eqn:Therm2}
    Q = C m \Delta T,
\end{equation}
where $C$ - specific heat of the material (thin film or substrate) and m - mass. In addition, thermal energy $Q$ is lost by the thermal conduction to the substrate and by convection to the environment (air). The amount of heat transferred from the thin film to the substrate can be described as 
\begin{equation}
    \label{eqn:Therm3}
    Q_c =  \frac{k_s S_s}{d_s} (T_f - T_s) \Delta t,
\end{equation}
where $k_s$ - thermal conductivity of the substrate, $S_s$ - contact surface area of the thin film with the substrate, $T_f$ and $T_s$ - temperatures of the film and substrate, respectively. The heat lost via convection to the environment is expressed as:
\begin{equation}
    \label{eqn:Therm4}
    Q_h = h S (T - T_0) \Delta t,
\end{equation}
where $h$ - convection coefficient, $S$ - surface area, $T$ and $T_0$ - temperatures of surface and environment.

The equation of the heat balance for such a system takes the following form:
\begin{equation}
 \begin{cases}
    Q_0 = Q_f + Q_c + Q_{hf}
\\[10pt]
    Q_c + Q_s + Q_{hs} = 0 
\end{cases}
\end{equation}

By combining expressions \eqref{eqn:Therm1}-\eqref{eqn:Therm4}, we obtain equation, which describes the time evolution of the temperature of the metal thin film:
\begin{equation}
    \label{eqn:ThermModel}
\renewcommand{\arraystretch}{25}
 \begin{cases}
	\dfrac{\Delta T_f}{\Delta t} + \left(\dfrac{k_s S_s}{m_f C_{pf} d_s}+\dfrac{h S_f}{m_f C_{pf}}\right) (T_f - T_0) - \dfrac{k_s S_s}{m_f C_{pf} d_s} (T_s - T_0) = \dfrac{\rho_f l_f I^2}{d_f a_f m_f C_{pf}}
\\[15pt]
	\dfrac{\Delta T_s}{\Delta t} + \left(\dfrac{k_s S_s}{m_s C_{ps} d_s}+\dfrac{h S_s}{m_s C_{ps}}\right) (T_f - T_0) - \dfrac{k_s S_s}{m_s C_{ps} d_s} (T_f - T_0) = 0.
\end{cases}
\end{equation}

Material parameters used for calculations are listed in Table \ref{tab:ThermMaterials}. The later stack consists of thick Si substrate capped with 100-nm-thick Bi thin film. Environment temperature is fixed at room temperature of $T_0 = 25^\circ$C. The convection coefficient us taken to be 30.

\begin{table}[h!]
\caption{Material parameters used for estimation of the charge current induced thermal effects in bismuth thin films on a silicon substrate.}
\label{tab:ThermMaterials}
\begin{tabular}{|c|c|ccc|ccc|}
\hline
          &          & \multicolumn{3}{c|}{Dimentions}                                       & \multicolumn{3}{c|}{Physical properties}                                                                               \\ \hline
          & Material & \multicolumn{1}{c|}{Length $l$}  & \multicolumn{1}{c|}{Width $a$} & Thickness $d$ & \multicolumn{1}{c|}{Density $\rho$}                      & \multicolumn{1}{c|}{Specific heat capacity $C_p$} & Thermal conductivity $k$ \\ \hline
Thin film & Bi       & \multicolumn{1}{c|}{$32\,\mu$m}    & \multicolumn{1}{c|}{$1.42\,\mu$m}  & $100\,$nm    & \multicolumn{1}{c|}{$9.78\,$g/cm$^3$} & \multicolumn{1}{c|}{$123.5\,$J/(kg K)}         & $7.97\,$W/(m K)         \\ \hline
Substrate & Si       & \multicolumn{1}{c|}{$14.5\,$mm} & \multicolumn{1}{c|}{$11\,$mm} & $0.5\,$mm    & \multicolumn{1}{c|}{$2.34\,$g/cm$^3$} & \multicolumn{1}{c|}{$711\,$J/(kg K)}           & $150.7\,$W/(m K)        \\ \hline
\end{tabular}
\end{table}

Solving Eq. \eqref{eqn:ThermModel} leads to the temporal dependence shown in Supplementary Fig.~\ref{fig:JouleHeating}. For the temperature induced effects to cause any harmonic distortions (generate parasitic AC signals), they have to take place at the timescale of half-period of the sourcing signal. In our experiments, we work with AC current at the fundamental frequency of $787\,$Hz with a period of $1.2\,$ms. At this timescale at maximum current amplitude of $90\,\mu$A, the expected thermally induced signal would be of about $300\,$pV in the longitudinal channel, which is below the measurement accuracy of our setup.

At the timescale of seconds,  the temperature difference induced by the Joule heating reaches $4.5\,$mK at $90\,\mu$A current amplitude. This effect will not induce any harmonic distortions but will cause constant change in the thin film temperature (similar to supplying DC current $I_\text{DC} = RMS(I_\text{amp})$). In the longitudinal channel this effect will cause nonlinearities at the fundamental harmonics of about $150\,$nV, which is $0.7\,$ppm of the signal measured of this channel. We consider this contribution to the total measured voltage to be negligible.

\begin{figure*}[h!]
	\begin{center}
		\includegraphics[width=0.65\textwidth]{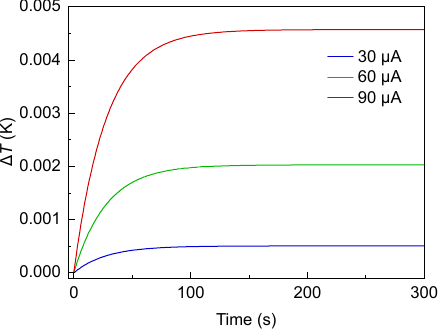}
	\end{center}
	\caption{\textbf{Joule heating in bismuth thin films on $0.5$-mm-thick silicon substrate.} Charge current induced heating of a $100$-nm-thick bismuth thin film structured into a $1.2\,\mu$m wide Hall cross structure. The data is shown for the case of dc current of different amplitudes flowing in the Hall cross. } 
	\label{fig:JouleHeating}
\end{figure*}

\clearpage

\section{Supplementary THz high harmonic generation measurements}

\subsection{THz fifth harmonic generation in Bi thin films}

\begin{figure*}[h!]
	\begin{center}
		\includegraphics[width=1\textwidth]{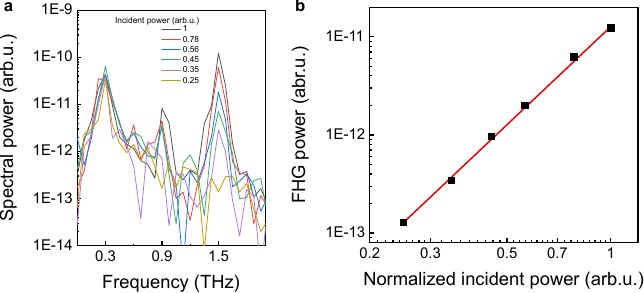}
	\end{center}
	\caption{\textbf{THz fifth harmonic generation (FGH) in Bi($100$\,nm)/Au($2$\,nm) thin films prepared on quartz glass}. (a) Power spectrum of the transmitted THz pulse at different incident power. The fundamental frequency of the incident light at $0.3$\,THz is attenuated by 5 orders of magnitude. Generation of the fifth harmonic is observed at $1.5$\,THz. (b) Spectral power of the generated fifth harmonic versus normalized power of the incident pulse.}
	\label{fig:FHG}
\end{figure*}

\clearpage

\subsection{THz third harmonic generation in Bi/Au heterostructures}

\begin{figure*}[h!]
	\begin{center}
		\includegraphics[width=0.75\textwidth]{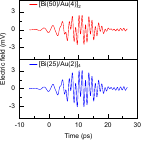}
	\end{center}
	\caption{\textbf{THz third harmonic generation (THG) in Bi/Au heterostructures (extended film) prepared on quartz glass.} Electric field profile in time domain of the transmitted THz radiation. We note that the light passed through a $300$\,GHz band-pass filter, which attenuated the fundamental harmonic by 5 orders of magnitude. 
 }
	\label{fig:Multilayers}
\end{figure*}

\clearpage

\subsection{Bismuth film thickness dependence of the THz third harmonic generation}

\begin{figure*}[h!]
	\begin{center}
		\includegraphics[width=0.75\textwidth]{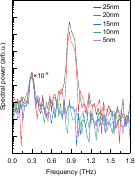}
	\end{center}
	\caption{\textbf{THz third harmonic generation (THG) in Bi($x\,$nm)/Au($2$\,nm) thin films prepard on quartz glass}. Transmitted spectral power measured after Bi thin films of different thickness $x$ varying from $5$ to $25$\,nm. Third harmonic generation is observed only in thin films thicker than $15$\,nm. 
 The absence of the THG in thinner bismuth films might be related to the island-like growth of the material or to the hybridization of the two surfaces of bismuth, opening the gap in the Rashba type surfaces - therefore quenching the THz THG.
The latter is an additional confirmation of the surface contribution to the measured THz third harmonics signal. }
	\label{fig:Thickness}
\end{figure*}

\clearpage

\subsection{Effect of thermal annealing on THz third harmonic generation}

\begin{figure*}[h!]
	\begin{center}
		\includegraphics[width=1\textwidth]{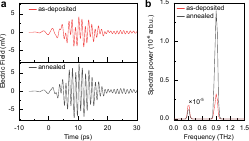}
	\end{center}
	\caption{\textbf{Effect of thermal annealing on THz third harmonic generation in Bi(100\,nm)/Au(2\,nm) thin films prepared on quartz glass.} (a) Transmitted THz signal in as-deposited (top panel) and thermally annealed in vacuum at $200^{\circ}$C for $3$\,h (bottom panel) thin film samples. (b) Power spectrum of the transmitted THz radiation. THz THG is enhanced in the sample after thermal annealing in vacuum. The observed enhancement of the efficiently of the THz THG is attributed to the enhancement of the quality of the ($0\,0\,1$) texture of the thin films. The fundamental THz signal ($\omega = 300\,$GHz) is attenuated by 5 orders of magnitude. 
 }
	\label{fig:Annealing}
\end{figure*}

\clearpage

\subsection{Optical second-harmonic generation at THz frequencies in curved devices}

We found conclusive evidence of the THz SHG in samples consisting of arrays of bismuth arcs. Due to small signal, the measurement even on arc arrays required the use of an accelerator TELBE at the HZDR. The resulting data are shown in Fig. 3 of main text. One can distinguish a clear second harmonic peak at $0.6~$THz (fundamental frequency: $0.3$\,THz). Importantly, we have also performed a comparative measurement with monolayer graphene using the same experimental setup [see Supplementary Fig.~\ref{fig:graphene}]. We clearly observe third and fifth harmonic generation whereas even harmonics are absent. This is completely compatible with the centrosymmetric nature of graphene that precludes a second-harmonic peak. 

We also observe a strong polarization dependence in the THz SHG signal of Bi arcs. As shown in Fig. 3 of the main text, the second-harmonic peak is enhanced when the incident electric field is polarized along the arc, \textit{i.e.} perpendicular to the symmetry axis of an individual arc structure. 
Based on this observation, we can argue that 
the geometry-induced nonlinear effects are much stronger than the intrinsic nonlinearities of the surface states.
The intrinsic contribution is expected to generate peaks of comparable power spectrum for the two orthogonal polarization because the longitudinal nonlinear conductivity component $\chi_{yyy}$ has magnitude equal to the transverse one $\chi_{yxx}$ [see Supplemental Theory Note E].
The strong polarization dependence therefore provides signature of the connection between the nonlinear transport regulated by the geometric NLHE and optical SHG at sub-millimeter frequencies.
 
Furthermore, we can parallel the presence of the geometric NLHE expected in various elemental materials with ubiquity of THz SHG in devices with curved stripes.
We performed additional high-harmonic generation measurements at TELBE accelerator using arrays of arcs made of 20-nm-thick gold. The results shown in Supplementary Fig.~\ref{fig:THzGold} prove that even though Au thin films do not have a visible THz THG, our arrays of arcs display a THz SHG peak at a much higher intensity level than the array of bismuth arcs (shown in Fig. 3(f) in the main text). Its features, namely the polarization dependence, is similar to elemental bismuth. 

\begin{figure*}[h!]
	\begin{center}
		\includegraphics[width=0.75\textwidth]{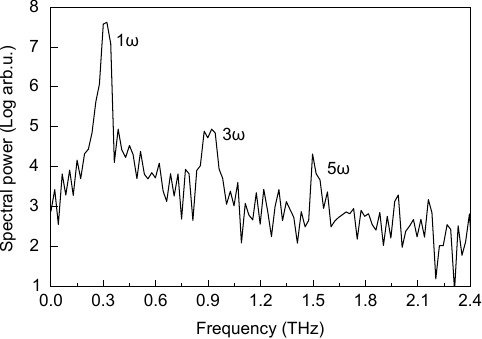}
	\end{center}
	\caption{\textbf{THz higher harmonic generation in a monolayer graphene.} Power spectrum of the THz radiation transmitted through a monolayer graphene transfered to quartz glass. Third and fifth harmonics are observed.} 
	\label{fig:graphene}
\end{figure*}

\begin{figure*}[h!]
	\begin{center}
		\includegraphics[width=\textwidth]{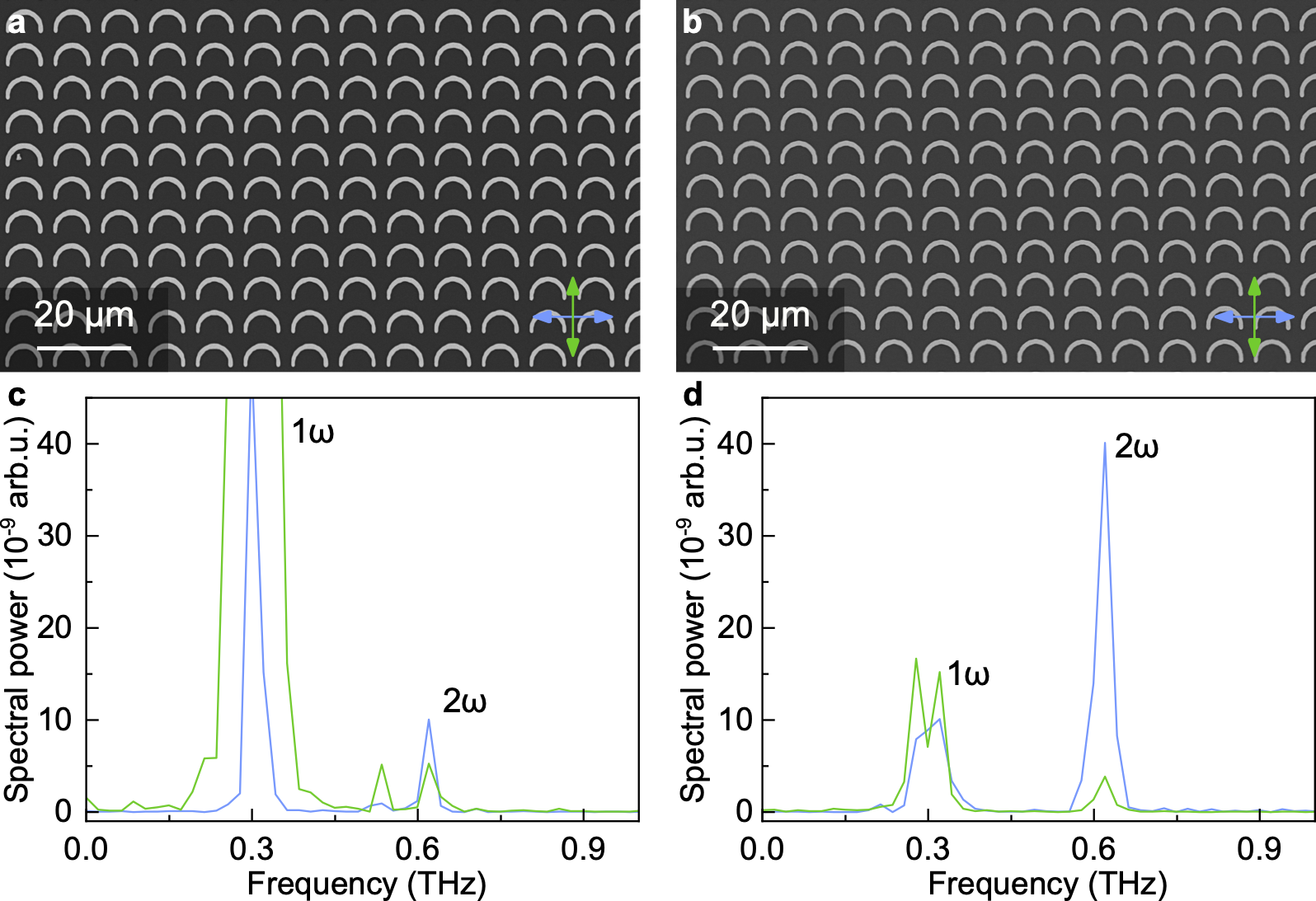}
	\end{center}
	\caption{\textbf{THz second harmonic generation (SHG) in arrays of arcs of 20-nm-thick Au prepared on a high purity silicon substrate.} (a,b) Scanning electron microscopy images of two samples of the arrays of arcs prepared using thin film deposition of $20$-nm-thick Au and optical lithography to realize arcs with inner radius of $3\,\mu$m and width of line of about $1\,\mu$m. Arrows in the bottom-right corner indicate polarization of the incident THz pulse. 
(c,d) Power spectrum of the transmitted THz radiation for the samples shown in panels (a) and (b), correspondingly (the data is shown in the same scaling as in Fig.3(f) of the main text). The intensity of the fundamental peak is suppressed by two band-pass filters. The background for each curve is also suppressed by subtracting measurements done in antiparallel polarization of THz pulse relative to the symmetry axis of the arcs, \textit{i.e.} $0-180^\circ$ and $90-270^\circ$. The color of lines in panels (c,d) corresponds to the color of arrows in panels (a,b) indicating the polarization of the incident light.}
	\label{fig:THzGold}
\end{figure*}

\clearpage

\section{Supplementary transport characterization}

\subsection{Harmonic transport measurement of a reference resistor}

\begin{figure*}[h!]
	\begin{center}
		\includegraphics[width=0.95\linewidth]{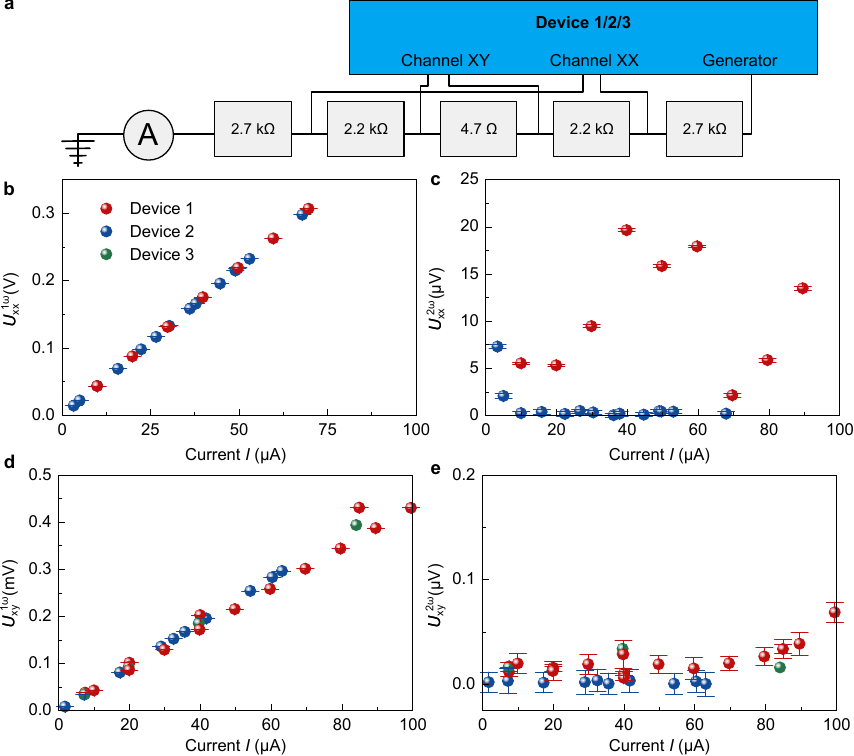}
	\end{center}
	\caption{\textbf{Harmonic transport measurement of a reference resistor.} 
 The reference measurement is performed of five resistors connected in series. Commercial surface mounted device (SMD) metal film resistors (Nova Electronic GmbH, Germany) were used for this reference measurement. The connection scheme of the resistors to the measurement device is shown in panel (a), A stands for ampermeter. This connection of 5 resistors resembles an electrical equivalent of the Hall cross structure upon measurement of both longitudinal and transverse voltage signals. The circuit mimics electrical properties of our $100$-nm-thick Bi-based Hall bar devices rendering similar current and input voltage ranges during the measurement. Device 1: lock-in amplifier HF2LI (Zurich Instruments AG, Switzerland). Device 2: Tensormeter RTM2 (Tensor Instruments, HZDR Innovation GmbH, Germany). Device 3: lock-in amplifier SR860 (Stanford Research Systems, USA). Longitudinal voltage measured at the (b) first harmonic and (c) second harmonic. The first harmonic signal shows Ohmic response to the supplied current. Transverse voltage measured at the (d) first harmonic and (e) second harmonic. The voltage measured in the Channel XY at the second harmonic $U_\text{xy}^{2\omega}$ is at the level of $20$\,nV. As nonlinear effects are not expected in reference resistors, we attribute this variation of the $U_\text{xy}^{2\omega}$ to the noise floor of the measurement setups. The measurements are done at the fundamental frequency of $787$\,Hz.}
	\label{fig:SI_refres}
\end{figure*}

\clearpage

\subsection{Harmonic transport measurement of Bi thin films on polymeric foils}

\begin{figure*}[h!]
	\begin{center}
		\includegraphics[width=0.95\linewidth]{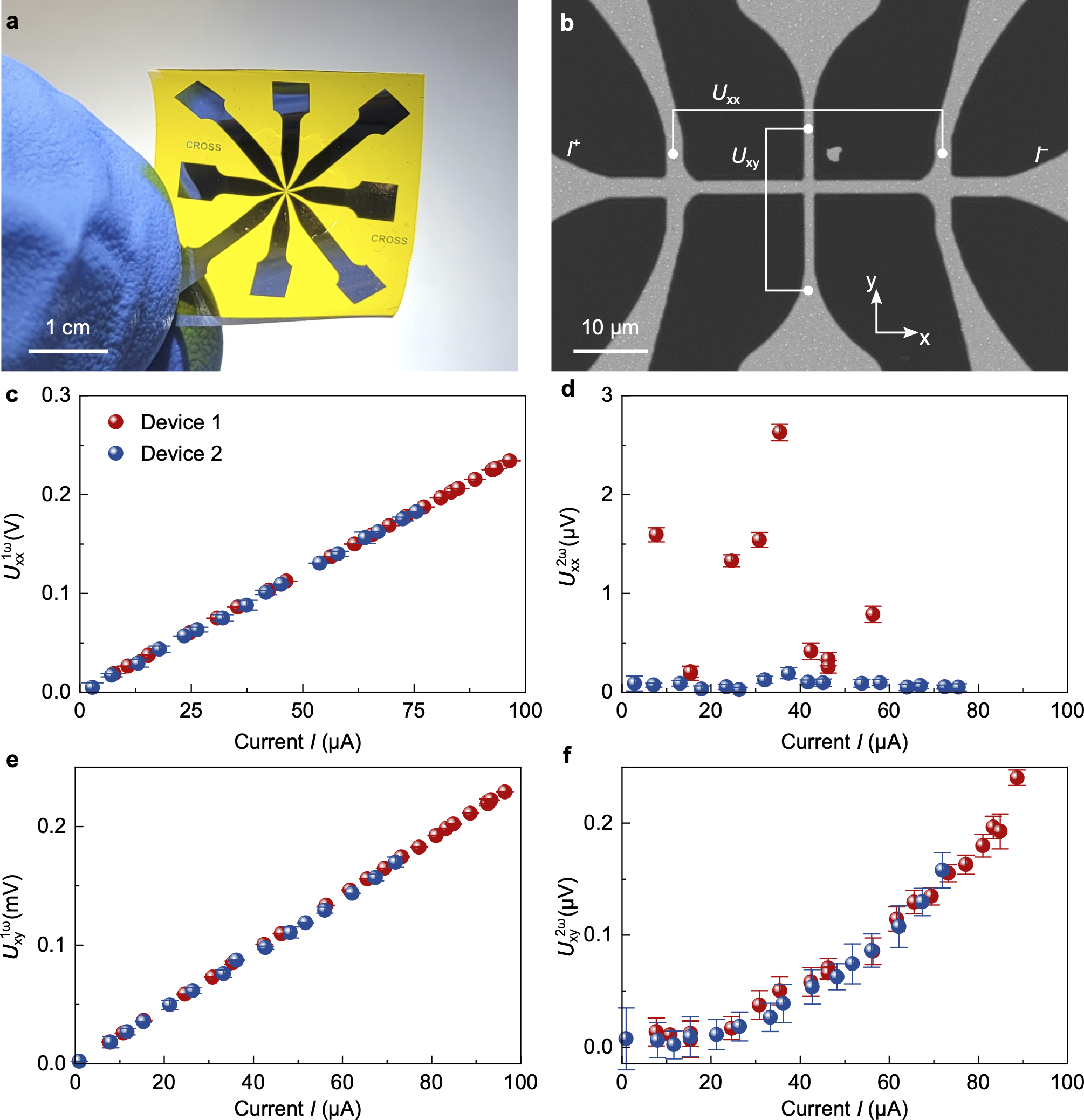}
	\end{center}
	\caption{\textbf{Harmonic measurement of a $100$-nm-thick Bi Hall cross prepared on a polymeric foil.} 
 (a) Photo of the $100$-nm-thick Bi thin film structured into a Hall cross on mechanically flexible $25$-$\mu$m-thick Kapton foil (DuPont, USA). The middle part of the device is covered with GE-varnish (Oxford Instruments, UK) to prevent bismuth thin film from oxidation. (b) Scanning electron microscopy image of the Hall cross structure. 
 Electrical measurements are done with 2 devices. 
  Device 1: lock-in amplifier HF2LI (Zurich Instruments AG, Switzerland). Device 2: Tensormeter RTM2 (Tensor Instruments, HZDR Innovation GmbH, Germany). Longitudinal voltage measured at the (c) first harmonic and (d) second harmonic. Transverse voltage measured at the (e) first harmonic and (f) second harmonic.  The measurements are done at the fundamental frequency of $787$\,Hz.
 }
 
	\label{fig:CrossKapton}
\end{figure*}

\clearpage

\subsection{Harmonic transport measurements of Bi Hall crosses using different devices}

\begin{figure*}[h!]
	\begin{center}
		\includegraphics[width=1\linewidth]{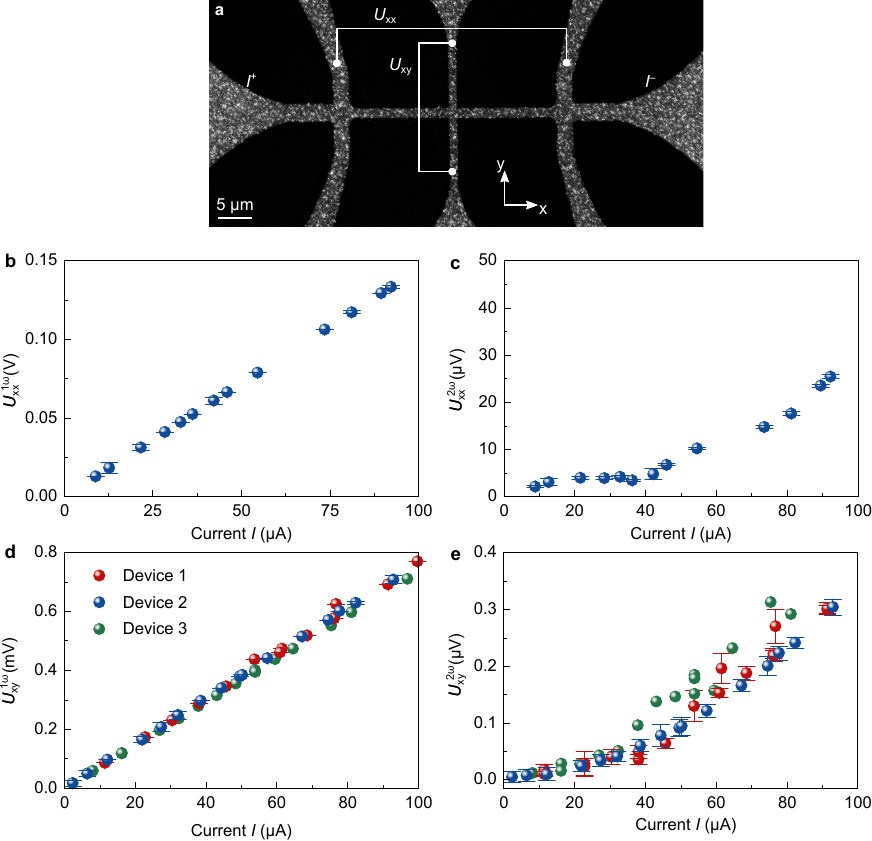}
	\end{center}
	\caption{\textbf{Harmonic measurement of a $100$-nm-thick Bi Hall cross prepared on Si/SiOx($500$\,nm).} 
 (a) Scanning electron microscopy image of the Hall cross structure. 
 Electrical measurements of the transverse voltage are done with 3 devices. 
  Device 1: lock-in amplifier HF2LI (Zurich Instruments AG, Switzerland). Device 2: Tensormeter RTM2 (Tensor Instruments, HZDR Innovation GmbH, Germany). Device 3: lock-in amplifier SR860 (Stanford Research Systems, USA). 
  Longitudinal voltage is measured using device 2 only. 
  Longitudinal voltage measured at the (b) first harmonic and (c) second harmonic. Transverse voltage measured at the (d) first harmonic and (e) second harmonic.  The measurements are done at the fundamental frequency of $787$\,Hz.
 }
	\label{fig:CrossSilicon}
\end{figure*}

\clearpage

\subsection{Harmonic transport measurements of arc-shaped Bi Hall bars using different devices}

\begin{figure*}[h!]
	\begin{center}
		\includegraphics[width=0.95\linewidth]{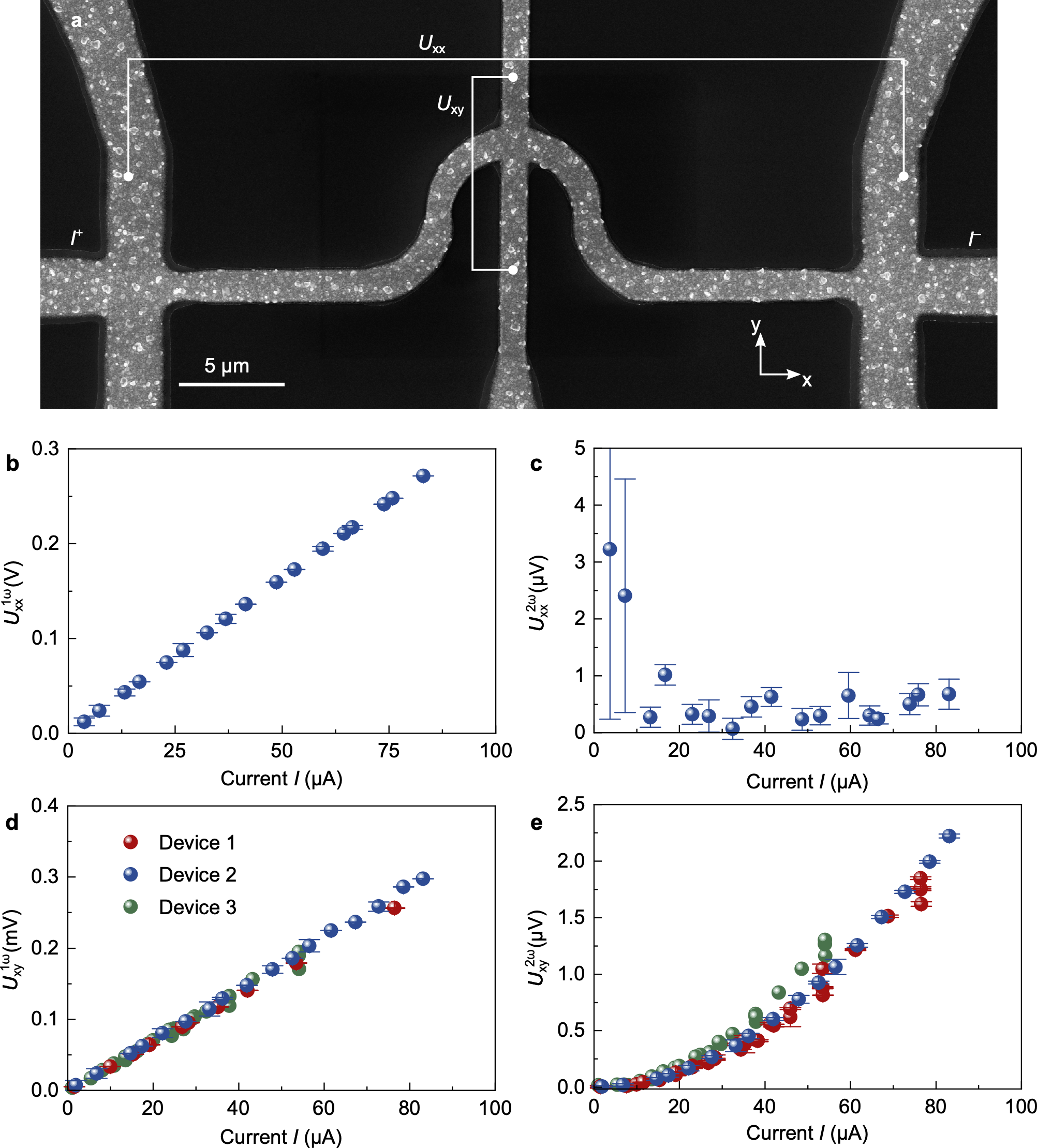}
	\end{center}
	\caption{\textbf{Harmonic measurement of an arc-shaped ($3\,\mu$m inner radius) $100$-nm-thick Bi Hall bar prepared on Si/SiOx($500$\,nm).} 
 (a) Scanning electron microscopy image of the arc-shaped Hall bar structure. 
 Electrical measurements of the transverse voltage are done with 3 devices. Longitudinal voltage is measured using device 2 only. 
  Device 1: lock-in amplifier HF2LI (Zurich Instruments AG, Switzerland). Device 2: Tensormeter RTM2 (Tensor Instruments, HZDR Innovation GmbH, Germany). Device 3: lock-in amplifier SR860 (Stanford Research Systems, USA). 
  Longitudinal voltage measured at the (b) first harmonic and (c) second harmonic. Transverse voltage measured at the (d) first harmonic and (e) second harmonic.  The measurements are done at the fundamental frequency of $787$\,Hz.
 }
	\label{fig:ArcSilicon}
\end{figure*}

\clearpage

\subsection{Harmonic transport measurements of Pt Hall crosses using different devices}

\begin{figure*}[h!]
	\begin{center}
		\includegraphics[width=1\linewidth]{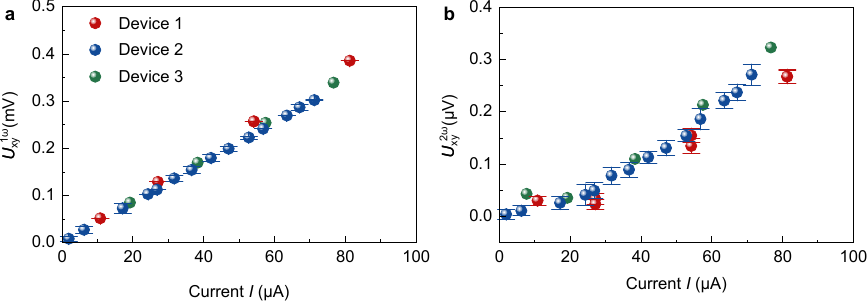}
	\end{center}
	\caption{\textbf{Harmonic measurement of a $5$-nm-thick Pt Hall cross prepared on Si/SiOx($500$\,nm).} 
 Electrical measurements of the transverse voltage are done with 3 devices. 
  Device 1: lock-in amplifier HF2LI (Zurich Instruments AG, Switzerland). Device 2: Tensormeter RTM2 (Tensor Instruments, HZDR Innovation GmbH, Germany). Device 3: lock-in amplifier SR860 (Stanford Research Systems, USA). 
  Transverse voltage measured at the (a) first harmonic and (b) second harmonic. Longitudinal voltage was not measured of this sample. The measurements are done at the fundamental frequency of $787$\,Hz.
 }
	\label{fig:PtHallCross}
\end{figure*}

\clearpage

\section{Fundamental achievements of this work and their potential technological implications}

We conclude by summarizing the fundamental advances connected with the demonstration of the nonlinear Hall effect at room temperature in bismuth. 
\begin{itemize}
    \item The nonlinear Hall effect of noncentrosymmetric quantum materials has been so far observed in materials with low-energy Dirac (transition metal dichalcogenides, graphene, topological insulator surface states) or Weyl (TaIrTe$_4$) quasiparticles. An exception is the electron gas formed at the (111) LaAlO$_3$/SrTiO$_3$ heterointerface, where, however, the nonlinear Hall effect survives up to $\simeq 30$K [see Ref.\,\cite{les22}]. Bismuth is the first material with conventional quasiparticles displaying room temperature nonlinear Hall effect. 
    
    \item Because of its bulk centrosymmetry, Bi realizes also the first example of a Berry-curvature-free conductive material displaying room-temperature nonlinear Hall effect due to the surface-induced electronic Berry curvature [see Ref.\,\cite{waw22}]. This paves the way to exploration of the nonlinear Hall effect in other noble metals, such as Pt, Rh or Ir. We note that we demonstrate that Pt Hall crosses also reveal nonlinear Hall response in transport (Supplementary Fig. \ref{fig:PtHallCross}).

    \item  Our study also shows the first example of nonlinear Hall effect in a material catalogued as an higher-order topological insulator. The connection between topology and Berry curvature-mediated transport effects has been demonstrated so far only in the more conventional (first-order) topological insulator Bi$_2$Se$_3$.

    \item Finally, our study also shows for the first time the realization of the geometric nonlinear Hall effect proposed in Ref.~\cite{sch19,gen22} using quasi-one-dimensional arc structures. In this regard, we would like to emphasize that in the semicircle structures studied in Ref.~\cite{sch19}, electron trajectories are not constrained to strictly follow circular paths of a given radius. 

\end{itemize}

As it concerns the technological relevance, it must be stressed that the nonlinear Hall effect of quantum materials bears potential for conversion of electromagnetic waves into direct current electricity. This process is a key ingredient for THz and mm-waves technologies. The rectification can be achieved with a diode combined with an antenna, called a rectenna. In conventional rectennas, the AC-DC power conversion efficiency abruptly drops to a few percent at 300 GHz. On the contrary, it has been recently shown [see Ref.~\cite{oni22}] that the power conversion efficiency for a nonlinear Hall rectifier can reach nearly 100$\%$. In this respect, the first demonstration of a room temperature nonlinear Hall effect in an elemental material fabricated with application-relevant thin film technology is of great importance for further device-related explorations.

\clearpage